\documentclass[useAMS,usenatbib,breaklinks=true]{mnras}
\usepackage{graphicx}
\usepackage{times}
\usepackage{amssymb}
\usepackage{amsmath}
\usepackage{hyperref}
\hypersetup{
    colorlinks=true,
    linkcolor=blue,
    citecolor=blue,
}

\title[SDSS DR16 LSS Catalogs] {The Completed SDSS-IV extended Baryon Oscillation Spectroscopic Survey: Large-scale Structure Catalogs for Cosmological Analysis}

\author[A.  J.  Ross et al.]{\parbox{\textwidth}{
Ashley J. Ross$^{1}$\thanks{Email: ross.1333@osu.edu},
Julian Bautista$^{2}$,
Rita Tojeiro$^{3}$,
Shadab Alam$^{4}$,
Stephen Bailey$^{5}$,
Etienne Burtin$^{6}$,
Johan Comparat$^{7}$,
Kyle S. Dawson$^{8}$,
Arnaud de Mattia$^{6}$,
H\'elion du Mas des Bourboux$^{8}$,
H\'ector Gil-Mar\'in$^{9,10}$,
Jiamin Hou$^{7}$,
Hui Kong$^{1}$,
Brad W. Lyke$^{11}$,
Faizan G. Mohammad$^{12,13}$,
John Moustakas$^{14}$,
Eva-Maria Mueller$^{2}$,
Adam D. Myers$^{11}$,
Will J. Percival$^{12,13,15}$,
Anand Raichoor$^{16}$,
Mehdi Rezaie$^{17}$,
Hee-Jong Seo$^{17}$,
Alex Smith$^{6}$,
Jeremy L. Tinker$^{18}$,
Pauline Zarrouk$^{19,6}$,
Cheng Zhao$^{16}$,
Gong-Bo Zhao$^{20,21}$,
Dmitry Bizyaev$^{22}$,
Jonathan Brinkmann$^{22}$,
Joel R. Brownstein$^{8}$,
Aurelio Carnero Rosell$^{23}$,
Sol\`ene Chabanier$^{6}$,
Peter D. Choi$^{24}$,
Chia-Hsun Chuang$^{25}$,
Irene Cruz-Gonzalez$^{26}$,
Axel de la Macorra$^{26}$,
Sylvain de la Torre$^{27}$,
Stephanie Escoffier$^{28}$,
Sebastien Fromenteau$^{29}$,
Alexandra Higley$^{11}$,
Eric Jullo$^{27}$,
Jean-Paul Kneib$^{16}$,
Jacob N. McLane$^{11}$,
Andrea Mu\~noz-Guti\'errez$^{26}$,
Richard Neveux$^{6}$,
Jeffrey A. Newman$^{30}$,
Christian Nitschelm$^{31}$,
Nathalie Palanque-Delabrouille$^{6}$,
Romain Paviot$^{27}$,
Anthony R. Pullen$^{18,32}$,
Graziano Rossi$^{24}$,
Vanina Ruhlmann-Kleider$^{6}$,
Donald P. Schneider$^{33,34}$,
Mariana Vargas Maga\~na$^{26}$,
M. Vivek$^{33,35}$,
Yucheng Zhang$^{18}$
 } \vspace*{4pt} \\ 
\scriptsize $^{1}$ Center for Cosmology and Astro-Particle Physics, Ohio State University, Columbus, Ohio, USA\vspace*{-2pt} \\ 
\scriptsize $^{2}$ Institute of Cosmology \& Gravitation, Dennis Sciama Building, University of Portsmouth, Portsmouth, PO1 3FX, UK\vspace*{-2pt} \\ 
\scriptsize $^{3}$ School of Physics and Astronomy, University of St Andrews, St Andrews, KY16 9SS, UK\vspace*{-2pt} \\ 
\scriptsize $^{4}$ Institute for Astronomy, University of Edinburgh, Royal Observatory, Edinburgh, EH9 3HJ, UK\vspace*{-2pt} \\ 
\scriptsize $^{5}$ Lawrence Berkeley National Laboratory, 1 Cyclotron Road, Berkeley, CA 94720, USA\vspace*{-2pt} \\ 
\scriptsize $^{6}$ IRFU,CEA, Universit\'e Paris-Saclay, F-91191 Gif-sur-Yvette, France\vspace*{-2pt} \\ 
\scriptsize $^{7}$ Max-Planck-Institut f\"ur Extraterrestrische Physik, Postfach 1312, Giessenbachstr., 85748 Garching bei M\"unchen, Germany\vspace*{-2pt} \\ 
\scriptsize $^{8}$ Department Physics and Astronomy, University of Utah, 115 S 1400 E, Salt Lake City, UT 84112, USA\vspace*{-2pt} \\ 
\scriptsize $^{9}$ Institut de Ci\`encies del Cosmos, Universitat de Barcelona, ICCUB, Mart\'i i Franqu\`es 1, E08028 Barcelona, Spain\vspace*{-2pt} \\ 
\scriptsize $^{10}$ Institut  d’Estudis  Espacials  de  Catalunya  (IEEC),  E08034  Barcelona,  Spain\vspace*{-2pt} \\ 
\scriptsize $^{11}$ Department of Physics and Astronomy, University of Wyoming, Laramie, WY 82071, USA\vspace*{-2pt} \\ 
\scriptsize $^{12}$ Waterloo Centre for Astrophysics, Department of Physics and Astronomy, University of Waterloo, Waterloo, ON N2L 3G1, Canada\vspace*{-2pt} \\ 
\scriptsize $^{13}$ Department of Physics and Astronomy, University of Waterloo, Waterloo, ON N2L 3G1, Canada\vspace*{-2pt} \\ 
\scriptsize $^{14}$ Department of Physics and Astronomy, Siena College, 515 Loudon Road, Loudonville, NY 12211, USA\vspace*{-2pt} \\ 
\scriptsize $^{15}$ Perimeter Institute for Theoretical Physics, 31 Caroline St. North, Waterloo, ON N2L 2Y5, Canada\vspace*{-2pt} \\ 
\scriptsize $^{16}$ Institute of Physics, Laboratory of Astrophysics, \'Ecole Polytechnique F\'ed\'erale de Lausanne (EPFL), Observatoire de Sauverny, 1290 Versoix, Switzerland\vspace*{-2pt} \\ 
\scriptsize $^{17}$ Department of Physics and Astronomy, Ohio University, 251B Clippinger Labs, Athens, OH 45701, USA\vspace*{-2pt} \\ 
\scriptsize $^{18}$ Center for Cosmology and Particle Physics, Department of Physics, New York University, New York, NY 10003, USA\vspace*{-2pt} \\ 
\scriptsize $^{19}$ Institute for Computational Cosmology, Dept. of Physics, University of Durham, South Road, Durham DH1 3LE, United Kingdom\vspace*{-2pt} \\ 
\scriptsize $^{20}$ National Astronomy Observatories, Chinese Academy of Science, Beijing, 100101, P.R. China\vspace*{-2pt} \\ 
\scriptsize $^{21}$ School of Astronomy and Space Science, University of Chinese Academy of Sciences, Beijing 100049, P.R.China\vspace*{-2pt} \\ 
\scriptsize $^{22}$ Apache Point Observatory and New Mexico State University, P.O. Box 59, Sunspot, NM 88349, USA\vspace*{-2pt} \\ 
\scriptsize $^{23}$ Centro de Investigaciones Energeticas, Medioambientales y Tecnologicas (CIEMAT), Madrid, Spain\vspace*{-2pt} \\ 
\scriptsize $^{24}$ Department of Physics and Astronomy, Sejong University, Seoul 143-747, Korea\vspace*{-2pt} \\ 
\scriptsize $^{25}$ Kavli Institute for Particle Astrophysics and Cosmology, Stanford University, 452 Lomita Mall, Stanford, CA 94305, USA\vspace*{-2pt} \\ 
\scriptsize $^{26}$ Instituto de F\'isica, Universidad Nacional Aut\'onoma de M\'exico, Apdo. Postal 20-364, Ciudad de M\'exico, M\'exico\vspace*{-2pt} \\ 
\scriptsize $^{27}$ Aix Marseille Universit\'e, CNRS, CNES, LAM, Marseille, France\vspace*{-2pt} \\ 
\scriptsize $^{28}$ CPPM, Aix Marseille Universit\'e, CNRS/IN2P3, Marseille, France\vspace*{-2pt} \\ 
\scriptsize $^{29}$ Instituto de Ciencias F\'isicas, Universidad Nacional Aut\'onoma de M\'exico, Av. Universidad s/n, 62210 Cuernavaca, Mor., M\'exico\vspace*{-2pt} \\ 
\scriptsize $^{30}$ PITT PACC, Department of Physics and Astronomy, University of Pittsburgh, 3941 O'Hara Street, Pittsburgh, PA 15260, USA\vspace*{-2pt} \\ 
\scriptsize $^{31}$ Centro de Astronom\'ia (CITEVA), Universidad de Antofagasta, Avenida Angamos 601, Antofagasta 1270300, Chile\vspace*{-2pt} \\ 
\scriptsize $^{32}$ Center for Computational Astrophysics, Flatiron Institute, New York, NY 10010, USA\vspace*{-2pt} \\ 
\scriptsize $^{33}$ Department of Astronomy and Astrophysics, The Pennsylvania State University, University Park, PA 16802, USA\vspace*{-2pt} \\ 
\scriptsize $^{34}$ Institute for Gravitation and the Cosmos, The Pennsylvania State University, University Park, PA 16802, USA\vspace*{-2pt} \\ 
\scriptsize $^{35}$ Indian Institute of Astrophysics, Koramangala, Bangalore 560034, India\vspace*{-2pt} \\ 
}

\date{} 

\pagerange{\pageref{firstpage}--\pageref{lastpage}} \pubyear{2019}

\begin{document}

\maketitle

\label{firstpage}

\begin{abstract}
We present large-scale structure catalogs from the completed extended Baryon Oscillation Spectroscopic Survey (eBOSS).
Derived from Sloan Digital Sky Survey (SDSS) -IV Data Release 16 (DR16), these catalogs provide the data samples, corrected for observational systematics, and random positions sampling the survey selection function. Combined, they allow large-scale clustering measurements suitable for testing cosmological models.
We describe the methods used to create these catalogs for the eBOSS DR16 Luminous Red Galaxy (LRG) and Quasar samples.
The quasar catalog contains 343,708 redshifts with $0.8 < z < 2.2$ over 4,808\,deg$^2$. 
We combine 174,816 eBOSS LRG redshifts over 4,242\,deg$^2$ in the redshift interval $0.6 < z < 1.0$ with SDSS-III BOSS LRGs
in the same redshift range to produce a combined sample of 377,458 galaxy redshifts distributed over 9,493\,deg$^2$. 
Improved algorithms for estimating redshifts allow that 98 per cent of LRG observations result in a successful redshift, with less than one per cent catastrophic failures ($\Delta z > 1000$ ${\rm km~s}^{-1}$). 
For quasars, these rates are 95 and 2 per cent (with $\Delta z > 3000$ ${\rm km~s}^{-1}$). 
We apply corrections for trends between the number densities of our samples and the properties of the imaging and spectroscopic data. For example, the quasar catalog obtains a $\chi^2$/DoF$= 776/10$ for a null test against imaging depth before corrections and a $\chi^2$/DoF$=6/8$ after. The catalogs, combined with careful consideration of the details of their construction found here-in, allow companion papers to present cosmological results with negligible impact from observational systematic uncertainties.

\end{abstract}


\section{Introduction}

The Sloan Digital Sky Surveys (SDSS) began in 1998. Since then, through phases I and II \citep{York00}, III \citep{Eis11}, and IV \citep{sdss4}, they have used the Sloan telescope \citep{Gunn06} in order to amass 2.6 million spectra of galaxies and quasars \citep{DR16}. The primary purpose of these observations that simultaneously place a single fiber on hundreds of extragalactic objects has been to create three dimensional maps of the structure of the Universe. From these maps, we observe the large-scale structure (LSS) of the Universe and thereby infer its bulk contents, dynamics, and structure formation history.  

During SDSS I and II, the measurement of the location of the baryon acoustic oscillation (BAO) feature in these maps was realized and developed as a robust and powerful method for obtaining geometrical measurements of the expansion history of the Universe and thus dark energy \citep{EH98,Eisenstein05,2dF,Percival10}. This motivated the Baryon Oscillation Spectroscopic Survey (BOSS; \citealt{Dawson12}) and extended BOSS (eBOSS; \citealt{eboss}) programs of SDSS-III and -IV. During these programs, considerable research was completed in order to use the signature of large-scale redshift-space distortions (RSD; \citealt{Kaiser87}) in the maps as a robust measure of the rate of structure formation (see \citealt{Acacia} and \citealt{DR16cosmo} for summaries of the developments), thereby allowing dynamical tests of dark energy and general relativity.

However, in order to confidently use these maps for these high-precision cosmological purposes, we must understand and account for how the survey design and operation (including all instrumental effects) impact the structure that we record. In essence, at every location in the observed space (angles and redshifts), we estimate the expected mean density (in the absence of any fluctuations due to clustering). This is commonly referred to as the survey `selection' or `window' function. It can be Poisson sampled by a set of random positions in the observed space (defined by the survey design and performance). Variations in the survey selection function can equally be accounted for by applying weights to either the data or random catalogs.

The complexity (in level of detail) in producing these matched data and random catalogs typically leads to independent public releases as SDSS value-added LSS catalog products, with publications describing their creation. For SDSS I and II, the details are in \cite{NYUVAGC}\footnote{The updated details for samples through DR7 are available at http://sdss.physics.nyu.edu/vagc/.}. For BOSS in SDSS-III, the details of catalogs extending to $z< 0.75$ are in \cite{Reid16}. SDSS-IV eBOSS completed on March 1st, 2019 and obtained four distinct samples for studies of large-scale clustering. Here, we describe the details of the creation of LSS catalogs for eBOSS quasars and luminous red galaxies (LRGs). Emission line galaxy (ELG) catalogs are described in \cite{RaichoorELGcat} and the Lyman-$\alpha$ forest analysis of high redshift quasars is described in \cite{dumasdesbourbouxDR16}.

The observed data and random catalogs we produce serve the primary purpose of obtaining BAO and RSD measurements from two-point statistics; i.e., the correlation function in configuration space and the power spectrum in Fourier space. The catalogs are, at their highest level, simply tables with one column for each of the three dimensions and extra columns that account for selection effects or provide weights that optimize these BAO and RSD analyses. The format is meant to allow efficient application of common correlation function and power spectrum estimators. While created to serve this particular purpose, the catalogs are documented and made public\footnote{https://data.sdss.org/sas/dr16/eboss/lss/catalogs/DR16/} in the hope that they will be useful for any LSS study.

For eBOSS, we calculated the catalogs using a development of the {\textsc MKSAMPLE} code, which traces its roots back to BOSS Data Release 9 \citep{DR9-BAO}, and is described in detail in \citet{Reid16}. In essence, {\textsc MKSAMPLE} was a framework for dealing with the particularities of the SDSS geometry, data model, and observing strategy. Very few of the original lines of code, or even algorithms themselves, are still used in the final eBOSS {\textsc MKALLSAMPLES} package. However, the underlying philosophy and basic set of necessary tasks remain almost the same. In this paper, we detail the changes and additions in the eBOSS process and describe the final catalogs that are produced.  

This paper is part of a series of papers presenting the completed eBOSS DR16 dataset and cosmological results derived from it, which are summarized in \cite{DR16cosmo}. The DR16 spectral reductions described in \cite{DR16} and the DR16 quasar catalog produced by \cite{LykeDR16} are vital inputs to the LSS catalogs we create. The LSS catalogs themselves were developed in close collaboration with the studies that obtain BAO and RSD results from the eBOSS DR16 data. For the LRGs, the correlation function is presented and used to measure BAO and RSD in \citet{BautistaDR16}, and the power spectrum in \citet{GilMarinDR16}. \citet{RossiDR16} presents the analysis of mock catalogs, designed to find and quantify any modeling systematic errors associated with the analysis of these data. The equivalent analyses of the quasar sample are presented in \citet{HouDR16}, \citet{NeveuxDR16} and \citet{SmithDR16}. The ELG catalogs are presented and analyzed in \citet{RaichoorELGcat}, further analyzed in \citet{TamoneDR16}, \citet{deMattiaELG}, and supported by simulations of the data presented in \citet{AlamDR16} and \citet{LinDR16}. Multi-tracer analysis utilizing the overlapping volume between the LRG and ELG samples is presented in \cite{WangMulti,ZhaoMulti}. The creation of approximate mocks to be used for covariance matrix estimation for all LSS samples is described in \cite{ZhaoDR16}. Finally the DR16 Lyman-$\alpha$ sample is presented and analyzed in \citet{dumasdesbourbouxDR16}. A summary of all SDSS BAO and RSD measurements with accompanying legacy
figures can be found here:
https://sdss.org/science/final-bao-and-rsd-measurements/ .  The full
cosmological interpretation of these measurements can be found here:
https://sdss.org/science/cosmology-results-from-eboss/ .

The types of target (quasar, LRG, ELG) are described in Section~\ref{sec:targ}, and the targeting criteria for each summarized. In Section~\ref{sec:obs}, we describe the eBOSS observing strategy.
The method used to measure redshifts is summarized in Section~\ref{sec:zed}, and the catalog creation in Section~\ref{sec:stats}. This section also includes details of how we have corrected for many observational effects including varying completeness, collision priority, close pairs, redshift failures, and systematic problems with the imaging data. This section ends with a review of the statistics for each sample, and provides details of how to use these catalogues.  A summary of the work is provided in Section~\ref{sec:conclusion}.\\

\section{eBOSS Targets}
\label{sec:targ}
eBOSS was designed to acquire redshifts for three types of tracers: quasars, LRGs, and ELGs. Each object selected from imaging data for follow-up spectroscopy is an eBOSS `target'. The selection criteria and motivation for these target samples are detailed elsewhere. Here, we record the essential details.

\subsection{Quasars and LRGs}

LSS quasar\footnote{In order to distinguish this work from the Lyman-$\alpha$ quasar sample, we will denote our sample as `LSS quasars'.} and LRG targets were selected using the same optical and infrared imaging data sets over the full SDSS imaging area.

The optical data were obtained during the SDSS-I/II \citep{York00}, and III \citep{Eis11} surveys using a drift-scanning mosaic CCD camera \citep{C} on the 2.5-meter Sloan Telescope \citep{Gunn06} at the Apache Point Observatory in New Mexico, USA. The five-passband ($u,g,r,i,z$; \citealt{F,Smith2002,Doi2010}) photometry was re-calibrated by \cite{Schlafly2012}, who applied the ``uber-calibration" technique presented in \cite{Pad08} to Pan-STARRS imaging \citep{Kaiser2010}. The photometry with updated calibrations was released with SDSS DR13 \citep{DR13} and was demonstrated to have sub-percent level residual calibration errors \citep{Finkbeiner2016}. This DR13 photometric data sample was used to inform the optical selection of eBOSS targets.

The infrared data were obtained using the Wide Field Infrared Survey Explorer (WISE, \citealt{Wright2010}). The WISE satellite observed the entire sky using four infrared channels centered at 3.4 $\mu$m (W1), 4.6 $\mu$m (W2), 12 $\mu$m (W3) and 22 $\mu$m (W4). We used the W1 and W2 data to identify eBOSS targets. All targeting is based on the publicly available {\it unWISE} coadded photometry, which obtained results for SDSS sources via `force-matching' \citep{Lang2014,Lang16}\footnote{These data have since been improved as described in \cite{Meisner19}.}.

The details of the quasar selection are presented in \cite{Myersquasartarg}, where it was demonstrated that SDSS+WISE imaging data can reliably select quasars with $0.9 < z < 2.2$. The method combined three essential pieces:
\begin{enumerate}
\item XDQSOz \citep{XDQSO} reporting a greater than 20 per cent chance of an object being a quasar at $z > 0.9$; 
\item an extinction corrected flux cut $g < 22 ~{\rm or}~ r < 22$; 
\item a mid-IR-optical color cut, which was proven to be efficient at removing stellar contaminants. 
\end{enumerate}
An important aspect of the quasar targets is that many were previously observed in SDSS I/II/III. Such targets are denoted as `legacy'; the LSS quasar legacy targets were not re-observed.  
Legacy targets are not isotropically distributed over the sky and thus must be treated carefully; these details are provided throughout Section \ref{sec:stats}. \cite{Ata} demonstrated that selecting quasars that were subsequently measured to have $0.8 < z < 2.2$ provided an excellent sample for LSS analyses. Here we will detail how we have built on these results to provide the final eBOSS quasar LSS catalogs. The target density is 112 deg$^{-2}$ within the 6,309 deg$^2$ area planned for eBOSS observation.

The full details of the LRG selection are given in \cite{Prakashtarg}. The goal was to obtain a sample at redshifts greater than the BOSS CMASS sample. In order to make it distinct, the sample was selected to be fainter in the $i$-band than BOSS CMASS galaxies \citep{Reid16}. Flux cuts were applied in the $i$- and $z$-bands in order to obtain targets bright enough to achieve a successful redshift. Optical/infrared color cuts achieved a sample with redshifts mostly greater than $z=0.6$, which is near where the density of the CMASS ceases to produce cosmic variance limited clustering measurements. \cite{BautistaDR14LRG} demonstrated the sample to be viable for LSS studies. The target density is 60 deg$^{-2}$ within the area planned for eBOSS observation. Here, we provide the details on the final LRG sample and combine it with the high redshift tail of the BOSS galaxy sample in order to provide one larger sample of LRGs with $z>0.6$.

Files containing the LRG and quasar target information applied to the full SDSS imaging were released\footnote{https://data.sdss.org/sas/dr14/eboss/target/ebosstarget/v0005/} in DR14 \citep{DR14}. They can be matched to the `full' files we describe later (Section \ref{sec:stats}) using the `OBJID\_TARGETING' column.

\subsection{ELGs}
The eBOSS ELG sample is unique from other eBOSS samples in that it does not use SDSS imaging for its target selection. Instead, ELG targets were selected from the DECam Legacy Survey (DECaLS; \citealt{DECaLS}) photometric catalog. The details of the selection are presented in \cite{RaichoorELGtarg}. There, it was demonstrated that within two separate $\sim$ 600~deg$^2$ regions, a $g$-band flux cut ($g < 22.9 (22.825)$ in the NGC (SGC) region) and a ($g-r$)/($r-z$) color selection were efficient at producing targets over the redshift range $0.6 < z < 1.1$ with sufficient $O[II]$ flux to obtain a good redshift. The target catalog was made public\footnote{https://data.sdss.org/sas/dr14/eboss/target/elg/decals/} in DR14. \cite{RaichoorELGcat} present further details on the ELG LSS catalog construction and its viability for LSS studies and we thus repeat few of them here.

\subsection{Other targets}

Observations of two additional samples of high redshift quasars for Lyman-$\alpha$ forest studies
\citep{Chabanier19,blomqvist19a,desaintagathe19a}
were also conducted during eBOSS. The first consisted of known $z>2.1$ quasars where increased signal-to-noise would lead to improved cosmology constraints. The second program used multi-epoch imaging data from the Palomar Transient Factory (PTF; \citealt{rau09,law09}) to select high-redshift quasar targets at a density
of 20 deg$^{-2}$ in regions with many epochs of photometry \citep{palanque-delabrouille16a}. These provide a random sampling of the foreground distribution of neutral
hydrogen and do not require a careful record of the selection function for cosmology studies.  
The Time Domain Spectroscopic Survey 
\citep[TDSS;][]{morganson15a} and the Spectroscopic Identification of eROSITA Sources
\citep[SPIDERS;][]{Clerc2016,Dwelly17} programs were also conducted simultaneously with eBOSS observations.
The Lyman-$\alpha$, TDSS, and SPIDERS samples do not directly
contribute to the clustering catalogs presented here, but the footprint of these observations is incorporated
into the clustering catalogs as will be described below.

\section{Spectroscopic Observing}
\label{sec:obs}

The eBOSS targets were primarily observed using the BOSS double-armed spectrographs \citep{Smee13} on the 2.5-meter Sloan Telescope \citep{Gunn06} at the Apache Point Observatory in New Mexico, USA. The exception is legacy quasar observations that used the original Sloan spectrograph. Here, we describe the details of the observational strategy and how it impacts our final sample, while defining key terminology. The details of how spectra are turned into redshift estimates are presented in Section \ref{sec:zed}.

Like BOSS, eBOSS observed 1000 targets at a time through fibers plugged into holes on pre-drilled aluminum plates.
Plates were placed at the focal plane of the telescope and the fibers fed directly into the two spectrographs.
On each plate, targets cannot be placed within 62$^{\prime\prime}$ of each other due to the physical size of the housing of the optical fiber \citep{Dawson12}. Targets are assigned to `collision groups' via a `Friends-of-Friends' algorithm with a 62$^{\prime\prime}$ linking length \citep{Reid16}. Any instance where a target is not observed because it is in a collision group is recorded as a `fiber collision'. A fraction of these collisions can be resolved in regions of overlapping plates. The `tiling' algorithm \citep{Blanton2003} determines the number of plates and the location of plate centers in celestial coordinates. In BOSS, tiling
over a fixed area produced a near-optimal solution of field locations that guaranteed 100\% completeness of non-collided targets for the primary clustering samples.  In eBOSS, the 100\% completeness requirement was relaxed for the LRG sample to increase the fiber efficiency and total survey area.  In both BOSS and eBOSS, each fixed area that was tiled in a single software run is referred to as a `chunk'.  The eBOSS LRG sample had a completeness of non-collided targets exceeding 95\% in
every relevant chunk of the survey.  Areas covered by a unique set of plates are `sectors'. Completeness statistics are determined on a per-sector basis.

LRG and LSS quasar targets were observed on the same plates, along with targets from the TDSS and SPIDERS programs.
TDSS and SPIDERS were each allocated an average surface density of targets of approximately 10 deg$^{-2}$.
These plates also contained fibers allocated to the two Lyman-$\alpha$ quasar target samples. It is possible for any of these samples to overlap in targeting with any other. Considering one particular sample, e.g., LSS quasars, the fact that it passed another sample's criteria is generally ignored; i.e., it is simply treated as any other LSS quasar when constructing the LSS quasar catalogs. When the target selection criteria are distinct, we must consider the effect of fiber collisions where LRG or quasar targets could not be allocated fibers because of these additional targets.   

Fiber collisions between different target categories were resolved based on the following priority: SPIDERS, TDSS, re-observation of known quasars, LSS quasars, and variability-selected quasars, with LRGs last. The fiber collision areas occupied by higher priority targets are treated in a `veto mask' that removes area from the window function of the desired clustering sample. Thus, these priorities result in an LRG sample that covers substantially less total area than the quasar sample, despite being observed at the same time across the same large-scale footprint. See Section \ref{sec:veto} for more details.

A significant portion of the LRG and quasar targets was observed in the Sloan Extended QUasar, ELG and LRG Survey (SEQUELS) that was designed as a pilot survey for eBOSS (see \citealt{eboss} for details). We treat SEQUELS targets that pass the eBOSS target selection the same as eBOSS observations, as the selection is a simple super-set of the ultimate selection. A list of the chunk numbers is given in Table~\ref{tab:chunktiles}. The SEQUELS targets are covered by chunks boss214 and boss217.

ELGs were observed separately from the quasars and LRGs in chunks eboss21, eboss22, eboss23, and eboss25. Some TDSS targets shared their plates and and were given equal priority. While they were observed in separate chunks, the ELG footprint overlaps with the LRG and quasar footprints and thus allows cross-correlation studies (e.g., \citealt{Alam19xcorr,WangMulti,ZhaoMulti}).

Although each chunk was tiled independently, there are some small regions of overlap in area. In other words, along some chunk edges, 
targets were assigned to more than one chunk. To account for this duplication, we removed the overlap area from the greater numbered chunk. Doing so provides a unique set of targets over the full eBOSS footprint. However, we always take the highest quality spectrum for duplicate tilings of the same target. For example, if a target received a fiber in both chunks eboss20 and eboss26 and the better spectrum was observed in eboss26, the redshift from the eboss26 observation is assigned. However, if the target in eboss20 was not assigned a fiber (e.g., due to a collision), but gets observed in eboss26, the eboss26 observation is not used in the clustering catalogs. This avoids biasing the selection probabilities within these regions. Such cases are fairly rare and are treated as if no fiber was placed on the target. The geometry file described in Section~\ref{sec:stats} cuts between chunks at the boundary of the highest-numbered chunk, corresponding to this selection.

In many cases, eBOSS chunks that were tiled did not have all of their plates observed. This is the primary source of incompleteness in the eBOSS catalogs. We denote the following classifications that lead to incompleteness for an eBOSS target within a tiled chunk:

\noindent\textbullet ~`close-pair': No fiber was placed on the target due to a fiber collision (un-resolved with overlapping plates) with a target of the same class.

\noindent\textbullet ~`missed': No fiber was placed on the target, not due to it being a close-pair. Observations will be missed primarily due to missing plates in overlap regions and can also occur when more than 1000 fibers would have been required to observe all targets in a given region.

\noindent\textbullet ~`wrong-chunk': A fiber was placed on the target only in the greater-numbered overlapping chunk.

\begin{figure}
\centering
\includegraphics[width=3.5in]{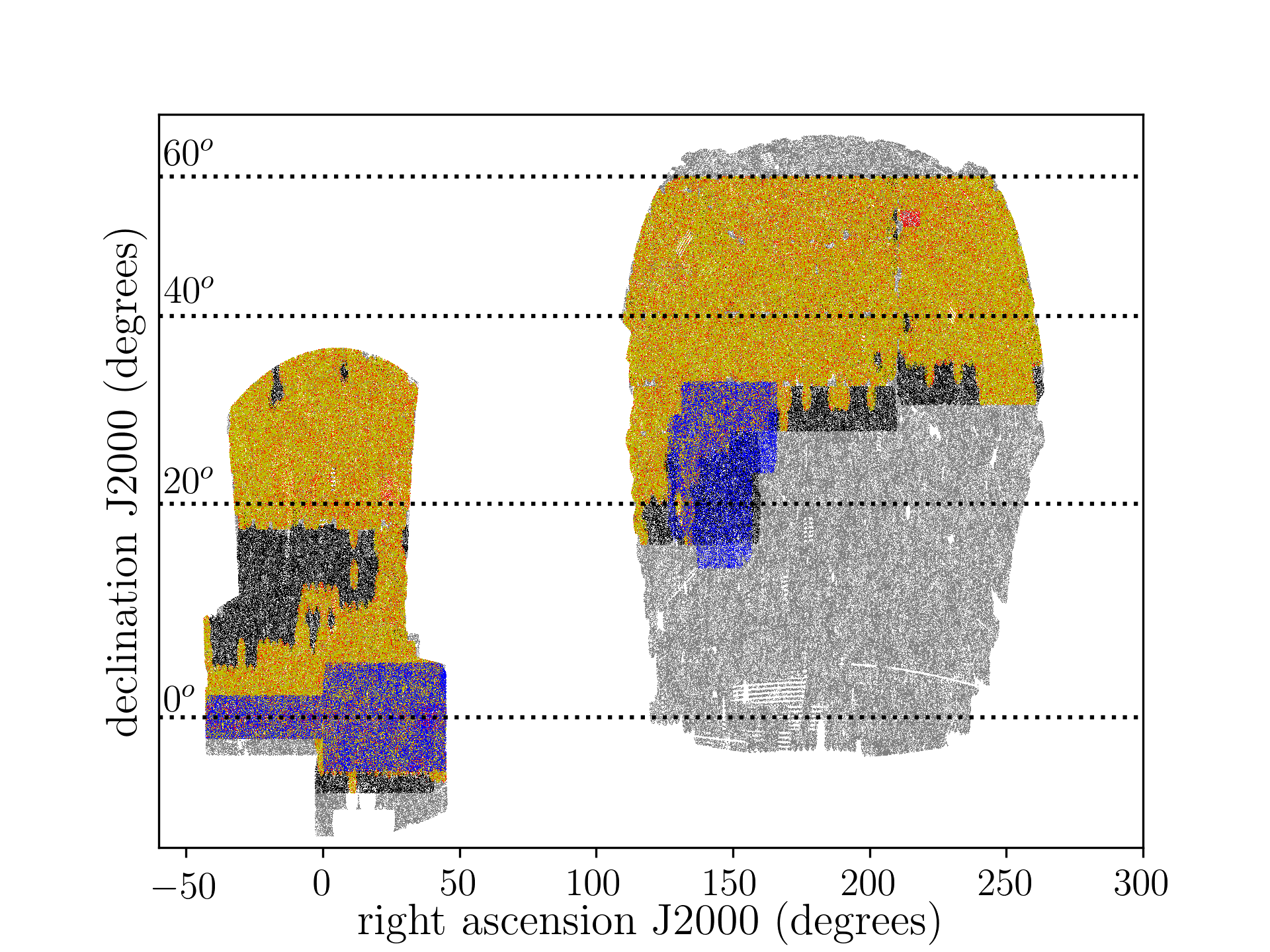}
  \caption{The footprint of eBOSS targets. Black points show LRG and quasar targets that were tiled but did not obtain spectroscopic observations (see text). Yellow points show quasars that were observed. They almost entirely overlap the red points, which show LRGs that were observed. Blue points show 20 per cent of the ELGs that were observed. The gray points are BOSS CMASS galaxies from their LSS catalogs. The CMASS data has their veto masks applied, while no such masks are applied for the eBOSS data in this plot. The eBOSS LRG and quasar footprints with veto masks applied are shown in Section \ref{sec:comp} and the equivalent ELG footprint is shown in \citet{RaichoorELGcat}. }
  \label{fig:DR16obs}
\end{figure}

Fig. \ref{fig:DR16obs} displays the sky positions of eBOSS targets that were tiled. The black points are LRG and quasar targets that were not observed. The colored points display eBOSS targets observed with plates determined to be `good'. The red points are LRG targets that were observed. They are mostly overlapped by the yellow points, which show the quasars that were observed. One can observe an area at ra, dec $\sim$ 225, 55 where there are no quasars. In this area, SDSS had previously obtained spectra for all of the quasar targets \citep[incorporated into the special Reverberation Mapping program;][]{2015ApJS..216....4S} and thus eBOSS did not re-observe them. The blue points display twenty per cent of the ELGs that were observed. This subsampling allows one to see the overlap with the LRG and quasar samples. The overlap of the ELG sample is complete in the South Galactic Cap (SGC; the filled area in the figure with right ascension $< 60$). In the North GC (NGC), the ELG data fully overlaps with the BOSS CMASS data, but the eBOSS LRG and quasar footprint only covers approximate half of the NGC ELG footprint.

\begin{table}
\centering
\caption{The eBOSS chunks, the number of tiles in the chunk, and the number of tiles with good observations in the chunk. (See text for definitions of `chunk'  and `tile'; see also Table 1 of \citealt{Reid16}.) boss214 and boss217 were SEQUELS chunks, while chunks 21, 22, 23 \& 25 were ELG plates. All others were standard LRG+QSO plates. SEQUELS tiled further chunks, which were unobserved at the end of the survey: the area in these chunks was recovered by eBOSS chunks, but without any overlap in the tiles. }
\begin{tabular}{lrr}
\hline
chunk	& \# of tiles & \# of good tiles\\\hline\hline
boss214 & 148 & 88 \\
boss217 & 74 & 29 \\
eboss1 & 199 & 195\\
eboss16 & 128 & 127\\
eboss2 & 98 &81\\
eboss20 & 42 & 42\\
eboss21 &46 & 46\\
eboss22 & 121 &121\\
eboss23 & 87 & 85\\
eboss24 & 81 &51\\
eboss25 &51 &51\\
eboss26 & 171 &76\\
eboss27 & 94 & 37\\
eboss3 & 204 & 180\\
eboss4 & 80 & 80\\
eboss5 & 70 & 70\\
eboss9 & 34 & 34\\

\hline\hline
\label{tab:chunktiles}
\end{tabular}
\end{table}

\section{Determining Redshifts}
\label{sec:zed}
The {\sc idlspec2d} spectral reduction pipeline reduces every eBOSS spectrum from a series of two-dimensional images
that span multiple exposures, to a single, wavelength-calibrated, one-dimensional spectrum.
The spectra that were used to generate the catalogs presented in this paper were processed with version
{\sc v5\_13\_0} of the data reduction pipeline.
This is the final version of the {\sc idlspec2d} software that will be used to process
clustering data obtained with the SDSS telescope.
A summary of the improvements to this software package over the course of eBOSS can be
found in the studies that first incorporated those improvements \citep{hutchinson16a,jensen16a,bautista17a}
and in the DR16 paper \citep{DR16}.

As in SDSS and BOSS, every spectrum is then assigned a classification of star, galaxy, or quasar, a redshift,
and a quality flag that indicates the robustness of the redshift estimate.  The redshift catalogs associated
with DR16 exactly follow the procedures described in \cite{DR13} and \cite{Bolton12}. 
However, different philosophies for redshift estimates and spectral classification were designed specifically for the eBOSS LSS catalogs. A new redshift estimate pipeline for galaxies was motivated by the challenges faced with the low signal-to-noise galaxy spectra. A new scheme that supplemented automated classifications with visual inspections was developed to characterize the very large number of quasar spectra obtained in eBOSS. We describe the new algorithms customized to LRG spectra in Section~\ref{subsec:lrgz} and briefly summarize the procedures for ELG and quasar spectra in Section~\ref{sec:qsoz} (these are described in greater detail in \citealt{RaichoorELGcat,LykeDR16}).

The relative success of
classification is divided into three cases: good redshift, redshift failure, and no chance of good redshift
(`bad fiber'). For all three LSS tracers, the bad fibers are determined based on the ZWARNING flag from the eBOSS
pipeline. Observations with bits 1 (`LITTLE COVERAGE'), 7 (`UNPLUGGED'), 8 (`BAD TARGET'), or 9 (`NO DATA')
had no chance of obtaining a good redshift and are classified as bad fibers.
As the cases of bad fibers are uncorrelated with the target properties, they are treated in the same manner as if they did not receive a fiber in the catalog creation, as described in Section \ref{sec:comp}. The following subsections detail how we classify between good redshifts and failures for LRGs and quasars. We describe the characterization of the spatial variation of redshift failures and our statistical corrections for them in Section~\ref{sec:specweight}.

\subsection{Redrock Redshift Estimates for the LRG Sample}\label{subsec:lrgz}
As discussed in \citet{eboss}, the spectra from the BOSS CMASS galaxy sample had sufficient signal-to-noise
to enable very reliable automated redshift classification using the same algorithms as those in the recently
released DR16 catalogs.
However, early in SDSS-IV, it became clear that these routines are not optimized for the fainter,
higher redshift LRG galaxies that comprise the eBOSS LRG sample.  When first applied
to the eBOSS samples, only about 70\% of the spectra were given good redshifts.
The high rate of redshift failures motivated the new development in the {\sc idlspec2d} spectral reduction pipeline
for higher quality one-dimensional spectra.  More significant improvements to the rate of good redshift estimation
were achieved through a new approach to redshift estimation.

The new redshift algorithm, {\sc redrock}\footnote{https://github.com/desihub/redrock; tagged version 0.14.0}, was developed for the Dark Energy Spectroscopic Instrument \citep[DESI;][]{DESI}. The {\sc redrock} team used an improved combination of the \cite{Bolton12} approach and an archetype \citep{Cool13} approach similar to that applied in {\sc redmonster} \citep{hutchinson16a}. Methods developed in \cite{Zhu16}\footnote{Parts of this associated code were used: https://github.com/guangtunbenzhu/SetCoverPy.} were incorporated in order to provide additional improvements. We describe the approach in more detail throughout the rest of this section.

The general process, which we expand on below, is as follows: Classification and redshift determination are performed via a fit of a linear combination of spectral templates to each spectrum.  Fitting is done over a range of redshifts for three different classes of templates that independently characterize stellar,
galaxy, and quasar spectral diversity.  Unlike the approach used in the BOSS redshift pipeline,
no nuisance terms are allowed to soak up flux calibration errors, intrinsic dust extinction, or other sources of spurious signal.
The redshift and spectral class that give the lowest
value of $\chi^2$ are considered the best description of the spectrum.
A fit is only considered reliable, or good, if it can be differentiated from the second best
fit by a sufficiently large difference in the $\chi^2$.  We denote this parameter as $\Delta \chi^2$.

The first improvement over the BOSS fitting routines
was the introduction of the instrument resolution to the spectral models.  Each model is generated
at a significantly higher resolution than offered by the BOSS spectrograph.  At each redshift, the
model is convolved with the wavelength-dependent estimate of the Gaussian profile that describes
the instrument resolution for that spectrum.  The inclusion of instrument performance in this
step allows better characterization of narrow spectral lines, particularly when there is a strong
variation in the resolution as a function of wavelength as often occurs near the detector edges.

The second improvement over the BOSS fitting routines is an introduction of new spectral templates
for galaxies and stars\footnote{https://github.com/desihub/redrock-templates; tagged version 2.6}.
Galaxy spectral templates are derived from a principal component analysis (PCA) decomposition
applied to a total of 20,000 theoretical galaxy spectra (Charlie Conroy 2014, private communication)
that span stellar age, metallicity,
and star formation rate\footnote{Specifically, these are broken up by DESI target class to have 10,000 ELGs, 5,000 LRGs, and 5,000 spectra representing the flux-limited `Bright Galaxy Sample'}.
Emission lines of varying equivalent width were
painted onto the theoretical galaxy spectra. The resulting PCA eigenspectra are therefore physically-motivated, as opposed
to the BOSS eigenspectra that were derived empirically from early data and are thus degraded from
noise and occasional spurious signal in the spectra. There are 10 galaxy PCA eigenspectra templates
that are used in linear combination to obtain redshifts for the entire eBOSS galaxy sample.  

The stellar templates were also derived
from a series of theoretical models divided approximately by stellar mass and evolutionary stage. The stellar templates were motivated by laboratory atomic data,
molecular data, and model atmospheres
(\citealt{Allende18},
Allende-Prieto et al. private communication).
A total of 30,000 template stars were used.
10,000 had spectral types A, B, F, G, K, or M. 20,000 white dwarf templates were used, split evenly between types DA and DB.
Broad TiO absorption features in
red dwarf spectra can masquerade as G-band or balmer breaks in high redshift galaxies. Thus,
extra care was taken to increase the diversity of M-type and K-type main-sequence stellar templates.
The introduction of these new templates was proven to reduce the rate of false
detections around $z=0.62$ and $z=1.02$. The CV-type stellar templates and the four quasar eigenspectra produced by \cite{Bolton12} and used in previous eBOSS analyses were copied into {\sc redrock}. The redshifts for the LSS quasar sample were determined determined as detailed in the following subsection (not by {\sc redrock}).

In the second element of the {\sc redrock} redshift classification scheme,
a subset of the spectral templates described above were used as archetype
models\footnote{https://github.com/desihub/redrock-archetypes; tagged
version 0.1}
to fit the spectra in a manner similar to {\sc redmonster}.
The motivation for this second step was to apply an additional filter on the spectral fitting
and exclude non-physical combinations of the eigenspectra that can generate erroneous redshift detections.
Archetype fitting was not performed over the full redshift range, but instead was performed only over the range of within 10,000 ${\rm km~s}^{-1}$ of the redshift estimate from a maximum of the three best-fit cases for each class (galaxy, quasar, star) from the first stage of classification.
For the redshift ranges where the spectral class was estimated to be a galaxy,
110 archetype galaxy templates were fit in combination with nuisance terms that control the amplitude
of the first three Legendre polynomials, meant to fit non-physical flux in the broadband spectrum.
Likewise, 40 stellar archetype spectra were fit to the spectrum for the redshifts where the PCA spectral
class was estimated to be stellar, and 64 archetypes were fit for class quasar.  The redshift and class that produced the lowest value of $\chi^2$
was then considered the best description of the spectrum.  The results from the archetype fits
superseded those from the PCA eigenspectra fitting and are used for the clustering catalogs.
The $\Delta \chi^2$ between the best two archetype fits was recorded and later used to define our redshift failure criteria.

To limit the number of interlopers in our LSS measurements, we established a requirement that limited
the number of misclassified, or `catastrophic failures', to be less than 1\%. 
A catastrophic failure for galaxies occurs when an object is confidently assigned a redshift that is in
error by more than 1000 km s$^{-1}$.
The final tuning to discriminate between good redshifts and redshift failures and to assess
the resulting rate of catastrophic failures was done empirically using multi-epoch spectra and sky spectra, as described below.

\begin{figure*}
\centering
 \includegraphics[width=3.25in]{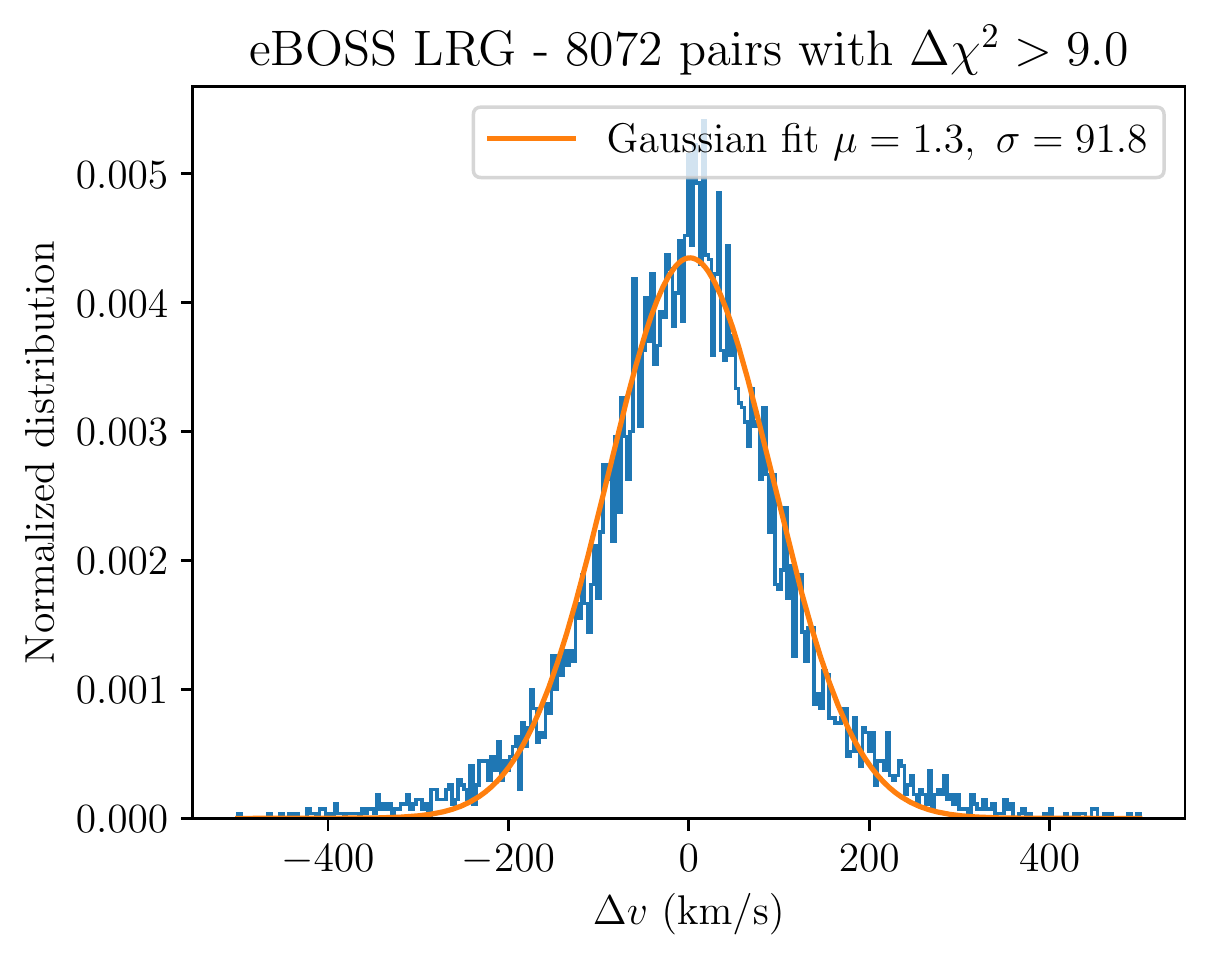}
 \includegraphics[width=3.25in]{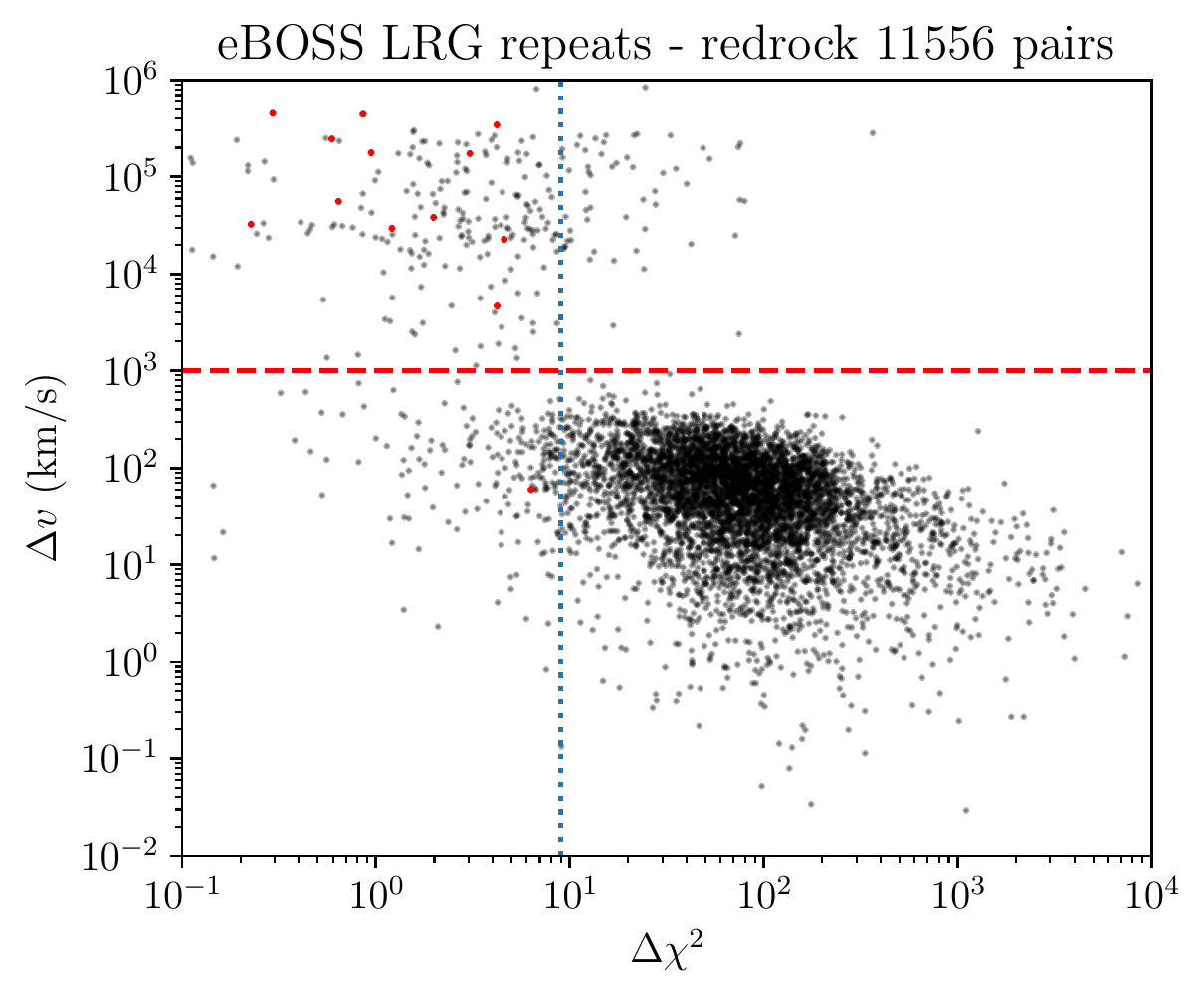}
  \caption{
{\bf Left:  }The distribution of $\Delta v$ over 8,071 pairs of observations of the same LRG target
at $0.6 < z < 1.0$ and with confident detections ($\Delta \chi^2 > 9$).
$\Delta v$ is the difference in
velocity between two redshift measurements of the same object.
The solid line shows the best-fit Gaussian model to the distribution
after requiring $|\Delta v| < 250$ km s$^{-1}$.  The mean and dispersion are shown in the legend.
{\bf Right:  }$\Delta v$ as a function of $\Delta \chi^2$, for 11,556 pairs where we apply no cut on redshift. $\Delta \chi^2$ represents the statistical
difference between the best and the second best fit spectral template to a single spectrum.  The lower $\Delta \chi^2$ in
the pair is considered the independent parameter.  Pairs in which one spectrum was fit with an archetype
spectral template with negative amplitude are presented in red. The horizontal red dashed line shows the limit of $\Delta v = 1000$ km/s above
which a pair is considered as a catastrophic failure. The dotted vertical line shows the
$\Delta \chi^2=9$, below which results are classified as redshift failures.
}
  \label{fig:zscatter}
\end{figure*}

We use a sample of multi-epoch spectra to identify a value of the $\Delta \chi^2$
that maximizes the number of good redshifts while maintaining sufficient purity in the catalog.
Many of these objects received more than one observation due to intentional reobservations of a
plate while others had multiple fiber assignments in the regions of plate overlap.  There were 11,556 pairs
of spectra used to perform this test.  For each pair of spectra, we determined the difference
in the redshift estimates, $\Delta v$.  The distribution of $\Delta v$ is shown in the
left panel of Fig.~\ref{fig:zscatter} while the results as a function of $\Delta \chi^2$ are presented
in the right panel.  Using the fit to the distribution, which we have cut to the $0.6 < z < 1.0$ redshift range used for the clustering catalogs, the mean redshift uncertainty for the LRG sample is 65.6 ${\rm km~s}^{-1}$ ($1/\sqrt{2}$ the width of the distribution in Fig. ~\ref{fig:zscatter}).  
An uncertainty of this scale is small compared to typical peculiar velocities and is thus absorbed into their modeling in the LSS analyses.

We then assessed the rate of catastrophic redshift failures by
counting the fraction of pairs that produced redshift estimates differing by more than 1000 ${\rm km~s}^{-1}$. 
For a threshold $\Delta \chi^2=9$ (rejecting 761 pairs), we find that 0.5\% of the 10,795 pairs produced a catastrophic redshift failure.
Under the assumption that one of the redshift estimates in the pair was correct, the resulting catastrophic
failure rate is estimated to be 0.25\%.

We then applied an additional level of filtering to further reduce the rate of catastrophic failures, which was to require a positive amplitude for the coefficient of the best-fitting archetype spectrum. In cases where the best fitting redshift was produced by an archetype template with a negative amplitude, we kept that redshift but set a flag indicating that the redshift was not to be trusted. These are counted as redshift failures in the down-stream analysis. We applied this condition based on tests of 365,243 sky-subtracted sky spectra. Without the requirement, we found that 10 per cent of these sky spectra were given a confident redshift estimate\footnote{These `redshifts' broadly sample the allowed redshift range with only minor structure
that appears to be caused by confusion from sky subtraction artifacts.} using the
$\Delta \chi^2=9$ threshold, whereas one would expect a negligible fraction of astrophysical spectra in those fibers. With the requirement, this was reduced to 4.4 per cent. While the positive-archetype requirement provides a significant improvement, this behavior indicates a systematic bias in the algorithm in the limit approaching zero signal.  We have not been able to identify the source of this bias. The spectra failing to meet the physicality condition are shown in red in Figure~\ref{fig:zscatter}.  One notes that
these pairs are most likely found at low values of $\Delta \chi^2$, as would be expected. There are only
0.04 per cent of LRG spectra in the full eBOSS sample that satisfy the $\Delta \chi^2=9$ condition but
fail to meet this threshold on positive archetype coefficients. Given the results on the sky spectra, we can expect a similar percentage of catastrophic failures in our LRG sample due to these false-positive confident redshifts; i.e., this is a negligibly small fraction. The requirement that the first coefficient be positive for the best-fitting archetype spectrum removes spurious detections from non-physical fits to the data, albeit at a very low rate.

After final classifications, the redshift completeness now approaches 98\% for the eBOSS LRG sample with a rate
of catastrophic failures estimated to be less than 1\%. 
These cases of catastrophic failures appear in the clustering catalogs without correction but are shown to
be sufficiently rare as to not bias the cosmological measurements.
Stars are a major contaminant, as they make up 9\% of the spectral classifications for the LRG sample.
An additional one per cent of spectra are classified as quasars and not used in the LSS catalogs.
In total, 88 per cent of LRG observations result in a good LRG redshift.

\subsection{ELG and Quasar Redshift Estimates}
\label{sec:qsoz}
We also utilize the {\sc redrock} code to make redshift estimates for the ELG spectroscopic sample.
The PCA and the archetype spectral templates are identical to those described above.
The requirement for $\Delta \chi^2$ between models
and the restriction on the archetype coefficients are also identical.
However, two additional criteria are applied to the ELG program:  the median signal-to-noise
per pixel must exceed $0.5$ in either the $i-$band or $z-$band region of the spectrum
and the measured continuum or
[OII] emission line strength must also pass the {\it a posteriori} flags defined and motivated in \citet{comparat16} and \citet{RaichoorELGtarg}; the criteria using these flags is ($zQ >=1$ or $zCont >= 2.5$).
The details of purity and completeness after each of these filters is presented in
\cite{RaichoorELGcat}. We are able to obtain secure redshifts for 91\% of ELG observations with
a catastrophic failure rate of less than 1\%.

We use a multi-stage process to determine the redshift and quality indicator for the quasar sample.
This process follows on the philosophy of \cite{paris18} and is fully described in \cite{LykeDR16}, which presents the `DR16Q' quasar catalog.
From these results, we used the following criteria to determine redshift failures: if an object was not
classified as a quasar by the automated decision-tree described in \cite{LykeDR16}\footnote{This classification is stored in the column named `MY\_CLASS\_PQN' in the `full' quasar catalog files.}
 and had an {\sc idlspec2d} {\sc ZWARNING} flag set
(not associated with the bad fibers described above), it was typed as a redshift failure. If no {\sc ZWARNING} flag was set, the observation was assigned the classification determined by the {\sc idlspec2d} pipeline.
Additionally, anything with a median signal-to-noise $<0.5$ per pixel across the spectrum was
classified as a redshift failure. All redshifts we use in the LSS catalogs were determined using the
{\sc redvsblue}\footnote{https://github.com/londumas/redvsblue} principle component analysis (PCA) algorithm described in \cite{LykeDR16} and stored in the \textsc{Z\_PCA} column within DR16Q. Within our redshift range of $0.8 < z < 2.2$, we find this redshift performs well both in terms of systematic and statistical uncertainties, as discussed below.\footnote{For the Lyman-$\alpha$ forest studies presented in \cite{dumasdesbourbouxDR16} \textsc{Z\_LYAWG} is used instead, where Lyman-$\alpha$ emission is masked. This is less of a concern in our redshift range.}
Visual inspection information is only used to evaluate the catastrophic failure rate, as described below.

The process results in 95 per cent of quasar target observations having a good redshift with a quasar, stellar, or galaxy classification.
Seven per cent of the observations are typed as galaxies and two per cent stars.
In total, 86 per cent of the eBOSS quasar observations are classified as having a good quasar
redshift.

The statistical uncertainties in quasar redshift estimates are computed empirically using repeat observations.
\cite{LykeDR16} find a typical statistical redshift error of 300 km s$^{-1}$ without strong redshift dependence. Systematic errors in redshift estimates are somewhat
more difficult to assess, as the emission lines that inform the fits are subject to internal dynamics and can be
shifted with respect to the quasar rest-frame. 
\cite{LykeDR16} study this by using results of repeat observations from the Reverberation Mapping program \citep{2015ApJS..216....4S} and find no evidence of a systematic uncertainty in the PCA redshifts with the range $0.8 < z < 2.2$; see their figure 3. 

Catastrophic failures are characterized via the 10,000 random visual inspections described in \cite{LykeDR16}.
From this set of 10,000, we select the LSS quasar targets that were classified as quasars and had eBOSS (not legacy) redshifts
$0.8 < z < 2.2$.  This sample provides a base set of 5,449 objects that we include as good quasar redshifts
in our clustering catalogs. Of these, the visual inspection found 1.2 per cent (63) were not quasars\footnote{No accurate new classification or redshift estimate was attempted but any resulting redshift would have been unlikely to be close to the original `quasar' redshift'}.
An additional 0.8 per cent (45) were determined to have redshifts with $\Delta v > 3000 $km s$^{-1}$
relative to the {\sc redvsblue} redshift. These combine for an estimated 2 per cent catastrophic failure rate on LSS quasar targets observed by eBOSS. All legacy redshifts had been visually inspected prior to eBOSS and determined to be good quasar redshifts. Legacy quasars make up 18 per cent of the quasar redshifts used in the LSS catalogs with $0.8 < z < 2.2$.
We thus estimate the total catastrophic failure rate to be 1.6 per cent for the LSS quasar sample (as the total fraction is 0.02$\times$0.82).

This statistical characterization of the distribution of redshift on uncertainties in our LSS quasar catalog is used in \cite{SmithDR16} to create simulations that are consistent with these results. Thus, \cite{HouDR16,NeveuxDR16} are able to determine the sensitivity of their BAO and RSD measurements to such redshift uncertainties and catastrophic failures and incorporate the results into their systematic error budgets.

\subsection{Redshift Distributions}

\begin{figure}
\centering
\includegraphics[width=3.5in]{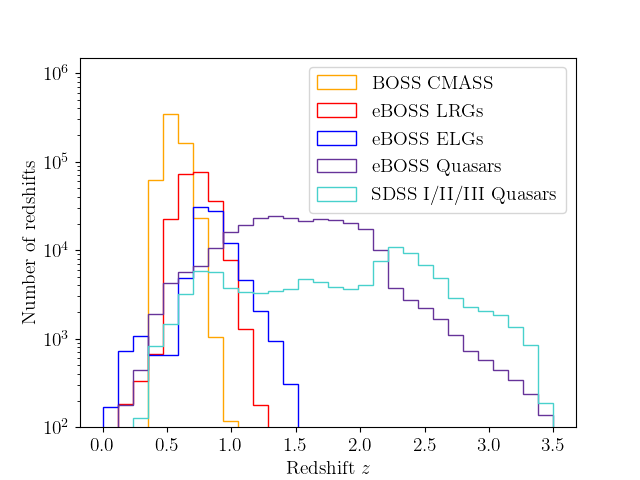}
  \caption{Histograms of the redshifts of samples used in eBOSS LSS analyses. The quasars are selected to pass our LSS sample target selection, as explained in the text, but many were already observed by previous generations of SDSS. The SDSS-III BOSS CMASS sample is included, as we combine this sample with eBOSS LRGs to produce one larger sample. }
  \label{fig:zDR16}
\end{figure}

Figure~\ref{fig:zDR16} displays redshifts from the samples used to create eBOSS LSS catalogs.
The SDSS I/II/III quasars were selected as eBOSS LSS quasar targets. These already had secure redshifts determined by visual inspection. Thus, for the LSS quasar sample, we use their previously observed spectra and
redshift estimates rather than re-observe them. See Section \ref{sec:targ} for more details. These legacy quasars span the whole redshift range and comprise approximately one quarter of the total quasar redshifts. The BOSS galaxies were not targeted by eBOSS, but we will use BOSS CMASS galaxies at $z > 0.6$ in order to create one combined sample of luminous galaxies with $z > 0.6$.

\section{Catalog creation}
\label{sec:stats}

In this section, we detail the catalog creation steps for the LRG and quasar samples. The steps are similar for the ELGs, but
those catalogs are described in \citet{RaichoorELGcat}. The order for catalog creation is:
\begin{itemize}
\item create randoms at constant surface density within the tiled footprint;
\item match between targets and spectroscopic observations; 
\item apply veto masks;
\item resolve fiber collisions and determine completeness;
\item assign weights to correct for fiber collisions and redshift failures;
\item cut on redshift and completeness;
\item assign weights that correct for systematic trends with foregrounds and imaging meta data;
\item assign redshift related information to random catalogs.
\end{itemize}

Many of these steps apply to both the data catalog and the random catalog that is used to quantify the window function.
The first operation is therefore to create a catalog with random angular positions at a density of 5000 deg$^{-2}$
within the geometry of the full tiled area, which is more than 40 times greater than the target density of the quasar sample (and more than 70 times greater than the LRG target density).  This area is a collation of the previously described chunks (with overlap removed)
and occupies 6309 deg$^2$. A polygon file that can be used with {\sc Mangle} \citep{Swanson2008} named `\texttt{eBOSS\_QSOandLRG\_fullfootprintgeometry\_noveto.ply}' defines this geometry, with a corresponding {\textsc FITS}\footnote{https://fits.gsfc.nasa.gov/fits\_standard.html} file that allows a mapping between polygons and sectors.

For each tracer, we release files containing the following types of tables:
\begin{itemize}
\item A table with a row for every unique target that was tiled and passes the veto masks; we denote these the `full' files. They contain all of the information on the target's photometry and spectra (if observed) and relevant IDs. One can use the column `OBJID\_TARGETING' to match to `objID' in the publicly available photometry\footnote{https://www.sdss.org/dr16/imaging/} and the columns `PLATE', `MJD', `FIBERID' to match to the publicly available spectra\footnote{https://www.sdss.org/dr16/spectro/spectro\_access/}.
\item A table with rows for only the data with good redshifts, with all mask, completeness, and redshift cuts applied. It contains only the columns that are necessary for calculating two-point statistics and matching to the full file. We denote these the `clustering' files.
\item A table of random points approximating the selection function of the clustering file for the data, to be processed in the same way as the data file for the calculation of two-point statistics.
\end{itemize}
The clustering files are produced separately for each Galactic hemisphere. 

\subsection{Matching Targets and Spectroscopic observations}

The galaxy catalog creation starts from the target sample. The information for all eBOSS targets within tiled chunks is collated. From this master list of eBOSS targets, the target sample in question is selected. Each target sample is then matched to spectroscopic observations. A first step is to cut the spectroscopic information to unique entries per target. For LRGs, this is done by selecting primary spectroscopic observations from the SDSS database (SPECPRIMARY = 1). For quasars, we use the DR16Q superset \citep{LykeDR16} catalog as the source of redshifts. The primary record (PRIM\_REC =1) is selected. 

For the quasars, we first match the legacy targets with their spectroscopic information. These objects were flagged in the target file as having already been observed and thus were removed from consideration by the tiling algorithm. We match these targets to DR16Q, populate the relevant spectral information (redshift, object type, etc.), and denote them as legacy. We then match the remaining targets based on their internal ID. For the LRGs, we go straight to matching based on internal ID.

After this matching, the following classifications are possible, which are stored as an integer value in the `IMATCH' column of the `full' file:
\begin{itemize}
\item a target can remain unobserved (IMATCH=0); denoted $_{\rm missed}$,
\item have a good eBOSS redshift that matches the targeted type (IMATCH=1; denoted $_{z,{\rm eboss}}$), 
\item have previously been determined to be a quasar with a good legacy redshift (IMATCH=2; denoted $_{\rm leg}$, relevant only for quasar targets), 
\item be a star (IMATCH=4; denoted $_{\rm star}$), 
\item be a redshift failure (IMATCH=7; denoted $_{\rm zfail}$, see Section \ref{sec:zed}), 
\item be identified as an object of the wrong target type (e.g., an LRG target is identified to be a quasar; IMATCH=9; denoted $_{\rm badclass}$), 
\item have previously been determined to be a legacy star (IMATCH=13; relevant only for quasar targets), 
\item be a bad fiber (IMATCH=14, see Section \ref{sec:zed}), 
\item or was not tiled in its target chunk (IMATCH=15, see Section \ref{sec:obs}). 
\end{itemize}
Objects with IMATCH=14,15 are treated the same as unobserved objects for calculating all subsequent statistics, i.e., we tabulate any quantity with the subscript $_{\rm missed}$ including the IMATCH=14 and 15 objects. Some IMATCH=2 objects will get re-assigned as IMATCH=8, based on completeness considerations, as described in Section 5.4. IMATCH 5 and 6 are not used.

\subsection{Veto Masks}
\label{sec:veto}

\begin{table*}
\centering
\caption{Statistics for the LRG and quasar samples within tiled areas and within the veto masks and completeness cuts we apply to obtain the final sample. Within each region, we list the number of targets, the number of good eBOSS (not legacy) redshifts, and its total area. Many of the veto masks have some overlap, but the statistics are presented individually (thus the total vetoed area is less the the sum of the area column). The statistics for the region with $C_{\rm eBOSS} \leq 0.5$ are presented after the veto masks have been applied and the statistics for $C_{ z} \leq 0.5$ are after the $C_{\rm eBOSS} \leq 0.5$ cut has been applied.}
\begin{tabular}{lrrrrrr}
\hline
Region & $N_{\rm LRG~tar}$ & $N_{\rm LRG~z}$  & $N_{\rm QSO~tar}$ & $N_{\rm QSO~z, eBOSS}$ & LRG Area (deg$^2$)  & QSO Area (deg$^2$)\\\hline\hline
Full Tiled Area & 377,633 & 230,935 & 703,521 & 340,386 & 6,309 & 6,309\\
veto masks:\\
LRG Collision Priority & 43,450 & 10,439 & - & - & 707 & - \\
QSO Collision Priority & - & - & 7,159 & 39 & -& 66 \\
Bad Field & 13,542 & 7,825 & 20,419 & 11,448  &  238 & 238 \\
Bright Star & 5,910 & 1,899 & 18,497 & 4,493  & 131 & 131 \\
Infrared Bright Star & 6,583 & 849 & - & -  & 72 & - \\
Bright Object & 1745 & 788 & 2,993 & 1212  & 28 & 28 \\
Centerpost & 36 & 0 & 204 & 0 & 0.6 & 0.6\\
Tiled Area, after veto mask & 311,848 & 209,894 & 655,521 & 325,226 & 5,223 & 5,858\\
Completeness cuts:\\
$C_{\rm eBOSS} \leq 0.5$ & 58,575 & 2,044 & 116,527 & 1179 & 978 & 1,047\\
$C_{ z} \leq 0.5$ & 53 & 20 & 366 & 32 & 1.6 & 3.4\\
Tiled area, after veto masks and completeness cuts& 253,220 & 207,830 & 538,628 & 324,015 & 4,242 & 4,808\\
\hline\hline
\label{tab:mask}
\end{tabular}
\end{table*}

After the matching and type assignment, a series of veto masks are applied to the targets and randoms. These masks and statistics describing what they remove are detailed in Table \ref{tab:mask}. Chunks covering a unique area of 6309 deg$^2$ were tiled for observation. Approximately 500 deg$^2$ of the area is vetoed from the quasar footprint and more than 1000 deg$^2$ is vetoed from the LRG footprint. Four veto masks are applied to each of the LRG and quasars. The bad field, bright star, and bright object masks are the same as applied to BOSS DR12 \citep{Reid16}. The centerpost mask removes the area at the center of the plate where no target can be observed (the centerpost pulls the center of the plate such that its curvature approximately matches the best-focus surface, see \citealt{Smee13}). 

The infrared bright star mask was applied to the LRG sample, as it was found that many spurious LRG targets exist around these stars. The size of the region that was masked is based on the WISE $W1$ magnitude, unless the 2MASS $K$-band magnitude was less than 2. Around each source, a circular region of 550$^{\prime\prime}$ was removed from consideration if either $W1$ or $K$ was less than 2 magnitudes. For fainter sources up to $W1=8$ we applied
\begin{equation}
r_{\rm IR mask} = (1397.5-569.34W1+79.88W1^2-3.75W1^3)^{\prime\prime}.
\end{equation}
Based on early data occupying 800 deg$^2$, 85 per cent of LRG targets removed by this mask were not LRGs. Thus, the mask was applied to LRG targets used for tiles eboss9 and greater so that the fibers could be assigned to targets more likely to produce good redshifts. These IR stars were not found to have any impact on quasar targets, beyond what is masked by the regular bright star mask.

The collision priority mask removes the 62$^{\prime\prime}$ radius area around where higher priority targets prevent any fiber to be assigned to the given target type. The LRGs had the lowest priority and the area of collision priority mask applied for them is thus nearly 700 deg$^2$. The overlapping plate geometry allows collisions between lower-priority LRGs and higher-priority targets to be resolved. However, these collisions are not fully resolved and some LRGs remain unobserved in these regions. We thus apply the conservative option of masking 62$^{\prime\prime}$ around every higher-priority target and accept losing the 10,439 good redshifts in this mask. 

Only TDSS and SPIDERS have greater priority than the quasars. Also, the quasar collisions in regions with overlapping plates are fully resolved. Thus, we only apply the quasar collision priority mask in single tile regions. This mask is only 66 deg$^2$ for the quasars and only 39 of the more than 7000 quasar targets removed by this mask have good redshifts; the number is greater than 0 only due to the fact that the center of the veto regions are within the single tile region but can extend out into the area with overlapping tiles. 

Note that by applying the veto masks to both galaxies and randoms, we are implicitly assuming that the regions removed are uncorrelated with the cosmological density fluctuations that we want to measure. This may be a slight concern where higher-priority targets overlap in redshift with the sample of interest. The main concern is clustering between $z<1$ quasars and our LRG sample. We apply no correction for this and expect it to be a minor effect on the LRG clustering given the substantially lower projected number densities of $z<1$ quasars compared to $0.6 < z < 1$ LRGs. This is not an issue for the ELG/LRG multi-tracer analysis, as these samples were observed on different plates and in different chunks, so the observation of one has no impact on the other.

In general, the veto masks were not applied to the target samples. Thus, many good redshifts were observed within these vetoed regions. For example, for the Bright Star and Bad Field masks, we do not trust that the photometry used to produce the target samples should produce isotropic samples suitable for large-scale structure. These areas thus tend to have proportionally fewer good redshifts. Across all veto masks, for LRGs, nine per cent of the good redshifts are vetoed, to be compared to 17 per cent of the tiled area. For the quasars, we lose 4.5 per cent of the good redshifts while removing 7.1 per cent of the tiled area.

\subsection{Spectroscopic Completeness Weights}
After the veto masks were applied, `close pairs', denoted $_{\rm cp}$, were assigned. Any object without a spectroscopic observation that shares a collision group with an object that obtained a spectroscopic observation is typed as a close pair and given IMATCH = 3. The distributions of these close pairs and also the redshift failures are not expected to be isotropic and close pairs are expected to be correlated with the density field itself. These sources of spectroscopic incompleteness require special treatment.

For the close pairs, the weights are assigned and equally distributed per collision group. All good observations in a collision group receive a weight that is 
\begin{equation}
w_{\rm cp} = \frac{N_{\rm cp} + N_{z, {\rm eboss}}+N_{\rm badclass}+N_{\rm star}}{N_{z, {\rm eboss}}+N_{\rm badclass}+N_{\rm star}},
\label{eq:wcp}
\end{equation}
where the $N$ are summed within each of these groups. Such a weighting provides unbiased transverse clustering on large-scales in configuration space. However, the radial clustering will be biased, and the issues are more severe in Fourier space \citep{HahnCP}. \cite{PIP} provide an unbiased solution for configuration space and \cite{PIPDR16} presents an application of these weights to eBOSS data. However, in the standard catalogs we simply provide $w_{\rm cp}$ and each individual analysis describes how the size of any remaining systematic biases how they are treated.

We provide corrections for redshift failures based on the spectrograph signal-to-noise in the $i$-band and the fiber ID. The likelihood of obtaining a good redshift naturally correlates with the signal-to-noise of the spectrum. The fiber ID correlates with the location of the spectrum on the CCD of the spectrograph, which in turn alters the signal-to-noise of the spectrum. The fiber ID also correlates with the expected location on the plate, resulting in large-scale signal-to-noise variations across the sky. 

We fit for trends between these quantities and the redshift efficiency, as defined below, and use the inverse of the trends as a weight. We define the number of good spectra associated with a particular
data subsample\footnote{A subsample can be, e.g., all spectra associated with a spectrograph on a single plate, all of the spectra associated with a given fiberID over all eBOSS observations, etc.} as 
\begin{equation}
N_{{\rm good} z} =  N_{z, {\rm eboss}}+N_{\rm badclass}+N_{\rm star}.
\label{eq:goodz}
\end{equation}
$N_{\rm all eBOSS}$ is then $N_{{\rm good} z} + N_{\rm zfail}$ and the redshift efficiency for any particular sub-sample is thus
\begin{equation}
f_{\rm good} = \frac{N_{{\rm good} z} }{N_{\rm all eBOSS}}.
\end{equation}
These statistics allow us to characterize the efficiency vs. particular aspects of the data (e.g., spectrograph, plate, fiberID) and derive statistical corrections for any trends that we find.

\begin{figure}
\centering
\includegraphics[width=3.5in]{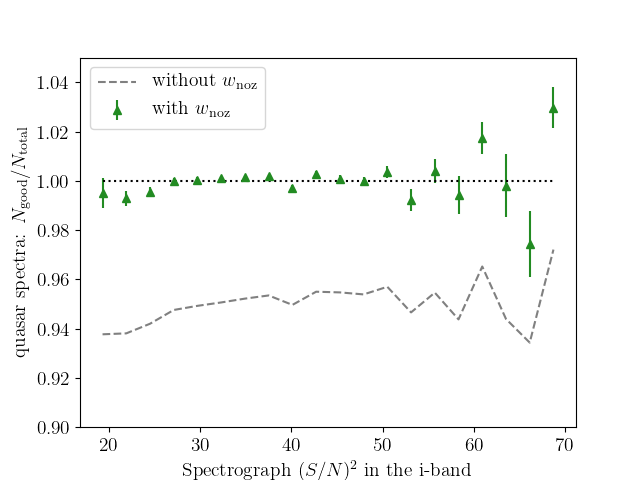}
\includegraphics[width=3.5in]{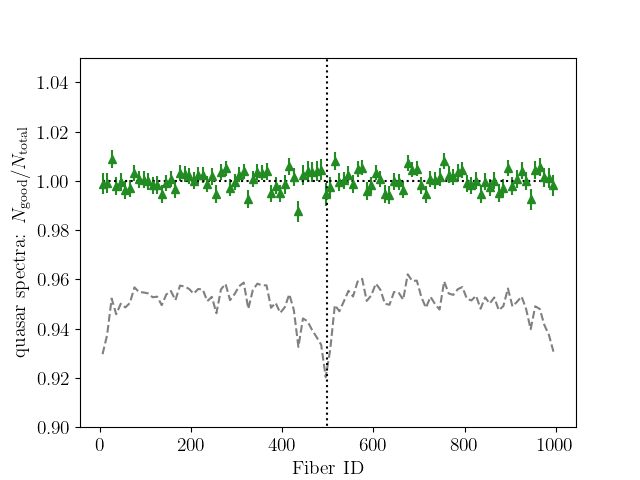}
  \caption{The fraction of good quasar spectra as function of the square of spectrograph signal-to-noise in the $i$-band ($S_i$ in text; top panel) and as a function of the fiber ID (bottom panel). The vertical dotted line at fiber ID 500 denotes the split between spectrographs 1 and 2.
 The gray dashed lines display the result when not applying the $w_{\rm noz}$ weights that we determine based on these quantities, as described in the text. }
  \label{fig:QSOzf}
\end{figure}

\begin{figure}
\centering
 \includegraphics[width=3.5in]{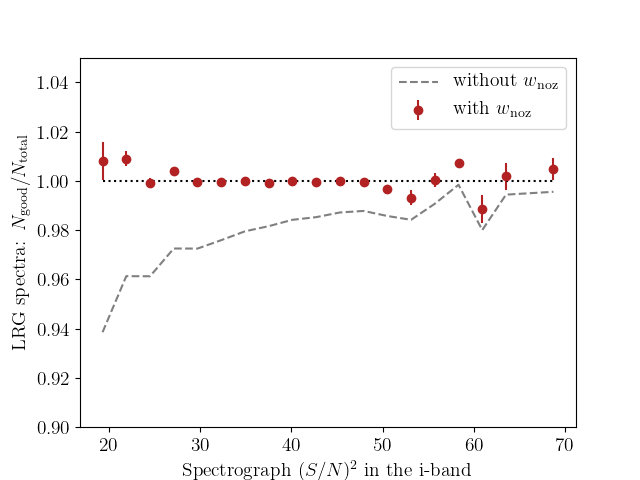}
\includegraphics[width=3.5in]{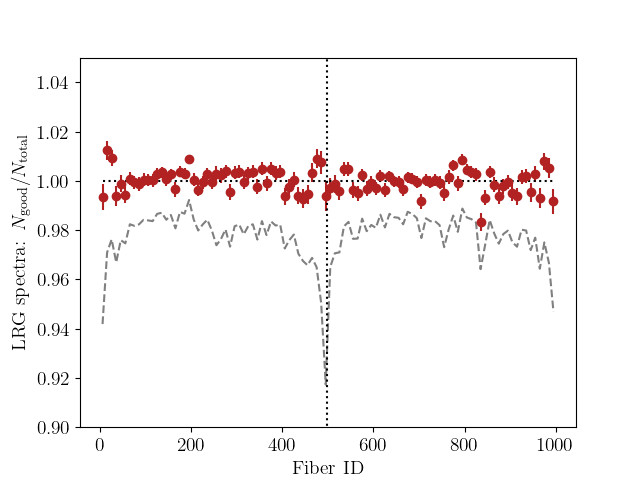}
  \caption{
  The fraction of good LRG spectra as function of the spectrograph signal to noise in the $i$-band (top panel)
and as a function of the fiber ID (bottom panel).
  The gray dashed lines display the result when not applying the $w_{\rm noz}$ weights that we determine based on these quantities, as described in the text.}
  \label{fig:LRGzf}
\end{figure}

We use the square of the spectrograph signal-to-noise, $S_i$, defined as the square of the median signal-to-noise of each spectral pixel in the $i$-band filter estimated at a photometric magnitude\footnote{This magnitude is motivated by the typical brightness of an LRG target.} $i=20.2$. This is empirically computed for each spectrograph independently using the combination of all measured
spectra from each observation. The trends in redshift efficiency vs. $S_i$ are shown in the bottom panels of Figs. \ref{fig:QSOzf} and \ref{fig:LRGzf}, where $f_{\rm good}$ is displayed as a function of the spectrograph signal-to-noise ratio in the $i$-band, $S_i$ using dashed gray curves. One can observe that the overall redshift efficiency is higher for the LRGs, but the trend with $S_i$ is stronger than it is for the quasars. In general, the LRG redshift efficiency is more dependent on the signal-to-noise level in a particular spectrum than is the quasar efficiency. 

The two classes of tracer have different dependencies on exposure depth due to the different spectral
features that inform the automated classification.  In the case of quasars, most targets have strong emission lines that
appear at a considerably higher signal-to-noise than does the continuum, thus facilitating relatively uniform
redshift efficiencies.  On the other hand, the primary features used for LRG spectral classification are absorption lines
that have a significance determined entirely by the signal-to-noise of the continuum.  Exposure depths were specifically tuned to
LRG redshift efficiencies, so variations in $S_i$ appear as variations in sensitivity to absorption line features and
variations in redshift efficiency.

We wish to (statistically) remove these trends from the data so that there is no spurious clustering signal in our catalogs that is associated with plate-to-plate variations in exposure depth. We apply the following steps, which mimic the modeling applied to the DR14 LRG sample in \cite{BautistaDR14LRG}. We find a linear relationship between $f_i = N_{{\rm good} z} /N_{\rm zfail}$ and $S_i$, which is determined per spectrograph per plate. One can observe that $f_{\rm good} = 1-1/(1+f_i)$. We perform the fits for each sample in each hemisphere. The fit is translated to a model for $f_{\rm good}$. Thus
\begin{equation}
f_{i,{\rm mod}} = a_S+b_SS_I
\end{equation}
and
\begin{equation}
f_{\rm good,mod} =  1-1/(1+a_S+b_SS_i).
\end{equation}
The inverse of this is used as a weight, $w_{{\rm noz,}S}$ to be applied to every good eBOSS observation.

The fiber ID, $r_{\rm ID}$, correlates directly with the location on the CCD where the spectrum is readout. The bottom panels of Figs. \ref{fig:QSOzf} and \ref{fig:LRGzf} use gray curves to display $f_{\rm good}$ as a function of $r_{\rm ID}$. Similar trends are observed in both samples, with the dependency again being stronger for the LRGs. The clearest trend is a decrease in redshift efficiency near $r_{\rm ID} = 1,500,1000$. These $r_{\rm ID}$ correspond to the edges of the CCDs, which are 1, 500 for spectrograph 1 and  501, 1000 for spectrograph 2. A good spectrum is more difficult to obtain near the edges of CCDs, as the optical quality decreases with increasing separation from the center of the CCD. The $r_{\rm ID}$ also correspond indirectly with the location on the focal plane, with low and high number occurring closer to the edge of the plates.
Some patterns are also observed near $r_{\rm ID}$ 250 and 750, where there are also small decreases in the redshift efficiency. These $r_{\rm ID}$ correspond to amplifier locations on the CCDs.

We assume a smooth model for the observed trend in redshift efficiencies with $r_{\rm ID}$, which is general enough to capture the trends described above. For each sample, hemisphere, and $r_{\rm ID}$ range 1-250,251-500,501-750,750-1000 we fit a relationship 
\begin{equation}
f_{\rm good,mod} =  A_r-B_r|r_{\rm ID}-C_r|^{D_r}.
\end{equation}
The inverse of this fit is used as a weight, $w_{{\rm noz,}r}$, to be applied to each good eBOSS observation. 

The corrections for the expected redshift failure rate per spectrograph, $w_{{\rm noz,}S}$, and per fiber, $w_{{\rm noz,}r}$ yield a combined weight:
\begin{equation}
w_{\rm noz} = w_{{\rm noz,}S}w_{{\rm noz,}r}.
\end{equation}
The weights are only applied to good eBOSS observations, i.e., objects that are not legacy and either have a good redshift or a securely determined alternative type (this is defined by Eq. \ref{eq:goodz}). We then normalize the weights such that their sum is equal to $1/f_{\rm good}$ for the full sample within a hemisphere. Thus, the points with error-bars in Figs. \ref{fig:QSOzf} and \ref{fig:LRGzf}, which display the results after applying the $w_{{\rm noz,}S}$, fluctuate around $f_{\rm good}=1$.

Figs. \ref{fig:QSOzf} and \ref{fig:LRGzf} show clear improvement in the trends with $S_i$ and $r_{\rm ID}$, with some residual scatter. The statistics should follow a binomial distribution, given that each observation can result in a success or failure. The uncertainty in each bin is thus $\sigma_{f {\rm bin}} = \left[N_{\rm all eBOSS,bin}f_{\rm zfail,bin}(1-f_{\rm zfail,bin})\right]^{1/2}/N_{\rm all eBOSS,bin}$. These error-bars are likely under-estimated. Random scatter exists in the signal-to-noise expected in each particular fiber, e.g., due to the fact that each will not have identical throughput. We expect these kinds of variation to be random with respect to position on the focal plane (and thus sky) and we do not attempt to model them. We use these uncertainty estimates to obtain $\chi^2$ values for the null test that $f_{\rm good}$ is constant with $S_i$ or $r_{\rm ID}$, but given the un-modeled sources of uncertainty and the fact that we do not attempt to account for any covariance between measurement bins, we do not expect $\chi^2$/dof = 1. The $\chi^2$ numbers are more useful in quantifying the degree of improvement. 

We use 20 bins to present the $S_i$ results. We find that the $\chi^2$ for the null test with $S_i$ for quasars decreases from 122 to 60 when the $w_{\rm noz}$ weights are applied. As one would expect, the improvement is more dramatic for the LRGs, where the $\chi^2$ decreases from 744 to 60. The scatter in the residuals, especially at high $S_i$, suggests that the uncertainties are under-estimated (as opposed to the high $\chi^2$ suggesting we have fit the wrong model).

We use 100 measurement bins to present the $r_{\rm ID}$ results. For the quasars, the $\chi^2$ compared to the null expectation improves from 404 to 114. For the LRGs, the $\chi^2$ improves from 761 to 224. There is a particularly strong outlier at $r_{\rm ID}\sim 850$. Some component of this high $\chi^2$ is likely due to under-estimation of the uncertainty. For both LRGs and quasars, the trends with $r_{\rm ID}$ are stronger than those with the physical focal plane positions and the correction for $r_{\rm ID}$ removes the trend observed in the $X$ focal plane position.

When applying both the close pair and redshift failure weights, we obtain an estimate of the number density as a function of redshift that we {\it would} have achieved from a complete spectroscopic survey that was able to extract good redshifts/reject bad targets with 100 per cent efficiency. The total spectroscopic completeness weight is given by
\begin{equation}
w_{\rm spec} = w_{\rm cp}w_{\rm noz}.
\end{equation}
{\it Note that this differs from previous BOSS and eBOSS analyses\footnote{These weights are included in the catalogs as `WEIGHT\_CP' and `WEIGHT\_NOZ'.}}, which defined the weights such that $w_{\rm spec,old} = w_{\rm cp,old}+w_{\rm noz,old}-1$ \citep{Reid16}.

\begin{table}
\centering
\caption{Basic properties of the quasar LSS catalogues, after veto masks. The quantities are first summed over the full tiled area, with no redshift or completeness cuts. The quantity $\rm N_{z,{\rm tot}}$ is the sum of the legacy and eBOSS redshifts. $\rm N_{eff}$ is the effective total number of quasars, after correcting for redshift failures and fibre-collisions, so it is the sum of all good eBOSS and legacy quasars, weighted by $w_{\rm cp}w_{\rm noz}$. These numbers are reported again after the completeness and then the redshift cuts that are both applied to produce the clustering catalogs.
 {\it Unweighted area} is the sum of the area of all sectors with $\rm C_{eBOSS} > 0.5$; {\it weighted area} multiplies this area by the completeness in each sector and {\em weighted area post-veto} multiplies this area by the total fraction of vetoed area. All other quantities are defined in the text. }
\begin{tabular}{lrrr}
\hline
	& SGC & NGC & Total \\\hline\hline
$ N_{ \rm eff}$ & 177,161 & 303,298 & 480,459 \\
$ N_{z,{\rm tot}}$ & 165,930 & 288,522 & 454,452 \\
$ N_{ z,{\rm eboss}}$ & 126,333 & 198,893 & 325,226 \\
$N_{z,{\rm legacy}}$ & 39,597 & 89,629 & 129,346 \\
$ N_{\rm zfail}$ & 8,162 & 10,616 & 18,778 \\
$ N_{\rm cp}$ & 4,832 & 6,878 & 11,710 \\
$ N_{\rm gal}$ & 9,386 & 18,655 & 28,041 \\
$ N_{\rm star}$ & 3,758 & 3,327 & 7,085 \\
$ N_{\rm star,leg}$ & 3,830 & 6,669 & 10,499 \\
after $C_{\rm eBOSS}>0.5$, $C_{z}>0.5$ cuts:\\
$ N_{\rm eff}$ & 176,080 & 302,306 & 478,386 \\
$ N_{z,{\rm tot}}$ & 164,929 & 287,602 & 452,531 \\
$N_{\rm eff}$, $0.8<z<2.2$  & 135,244 & 231,183 & 366,427 \\
$ N_{z,{\rm tot}}$, $0.8<z<2.2$ & 125,499 & 218,209 & 343,708 \\
Area (deg$^2$) & 1,884 & 2,924  & 4,808 \\
Weighted area (deg$^2$) & 1,839 & 2,860 & 4,699\\
\hline\hline
\label{tab:QSOstat}
\end{tabular}
\end{table}

\begin{table}
\centering
\caption{Basic properties of the LRG LSS catalogues, after veto masks. Quantities are the same as those defined in Table \ref{tab:QSOstat} (with $ {N}_{\rm QSO}$ the equivalent of $ N_{\rm gal}$).}
\begin{tabular}{lrrr}
\hline
	& SGC & NGC & Total \\\hline\hline
$ {N}_{\rm eff}$ & 87,607 & 134,695 & 222,302 \\
$ {N}_{z,{\rm tot}}$ & 82,607 & 127,287 & 209,894 \\
$ {N}_{\rm zfail}$ & 2,205 & 3,019 & 5,224 \\
$ {N}_{\rm cp}$ & 3,436 & 4,950 & 8,386 \\
$ {N}_{\rm QSO}$ & 1,254 & 1,635 & 2,889  \\
$ {N}_{\rm star}$ & 10,749 & 10,017 & 20,766 \\
after $C_{\rm eBOSS}>0.5$, $C_{z}>0.5$ cuts:\\
$ {N}_{\rm eff}$ & 86,511 & 133,540 & 220,051 \\
$ {N}_{z,{\rm tot}}$ & 81,600 & 126,230 & 207,830 \\
$ {N}_{\rm eff}$, $0.6<z<1.0$  & 71,427 & 113,868 & 185,295 \\
$ {N}_{z,{\rm tot}}$, $0.6<z<1.0$  & 67,316 & 107,500 & 174,816 \\
Area  post-veto (deg$^2$) & 1,676 & 2,566  & 4,242 \\
Weighted area post-veto (deg$^2$) & 1,627 & 2,476 & 4,103\\
\hline\hline
\label{tab:LRGstat}
\end{tabular}
\end{table}
  \label{sec:specweight}

\subsection{Completeness}
\label{sec:comp}
We determine the completeness per sector (each area covered by a unique set of plates) in the same manner as previous BOSS and eBOSS studies:
\begin{equation}
C_{\rm eBOSS} = \frac{N_{\rm z,eboss}+N_{\rm cp} +N_{\rm badclass}+N_{\rm star} + N_{\rm zfail}}{N_{\rm z,eboss}+N_{\rm cp} +N_{\rm badclass}+N_{\rm star} + N_{\rm zfail}+N_{\rm missed}}.
\end{equation}

\noindent This is basically everything that had a chance of providing a good spectrum, plus close pairs divided by the same plus the remaining number of targets in the sector that were not legacy. This provides us with an angular completeness that is not tied to the instrumental performance (like the redshift failure rate) or the small-scale clustering of the sample in question (like the close-pair weights are). The fluctuations in $C_{\rm eBOSS}$ can thus be treated in the random catalogs. {\it Unlike} previous BOSS and eBOSS analyses, we do not subsample the random catalog based on this completeness. Rather, it is used as a weight for all relevant calculations. This has the same effect, with slightly better noise properties. The primary advantage is that one can ignore $C_{\rm eBOSS}$ and still calculate any angular statistics on the sample, without any regard to the spectroscopic completeness. 

After determining $C_{\rm eBOSS}$ for eBOSS quasars, we use it to sub-sample the legacy observations. This only applies to the quasar sample. The legacy observations represent all quasar targets that had existing SDSS spectra (from any of SDSS I,II, or III). They were removed from the target list sent to the tiling algorithm. Thus, to be in the legacy sample an object must have already been observed and our legacy sample is by definition complete. However, we are weighting the randoms by $C_{\rm eBOSS}$ and these randoms are meant to be compared to the full eBOSS quasar sample (including legacy). For this to work, we must artificially impose $C_{\rm eBOSS}$ on the legacy sample. To accomplish this, we apply the same choice as previous BOSS and eBOSS analyses: we discard a fraction $1-C_{\rm eBOSS}$ of legacy observations from every sector. In this way, if we now include the remaining legacy observations in $N_{\rm z,eboss}$ and the discarded ones in $N_{\rm missed}$, we will recover the same $C_{\rm eBOSS}$ as originally determined (and as imparted into the randoms). The discarded objects are assigned IMATCH=8 and included in the `full' catalogs, but are otherwise ignored in the subsequent analyses. This choice allows us to avoid having to include the spatial distribution of legacy observations in the sample mask. From this point onward, legacy and eBOSS redshifts are treated in exactly the same way.  

\begin{figure}
\centering
  \includegraphics[width=3.25in]{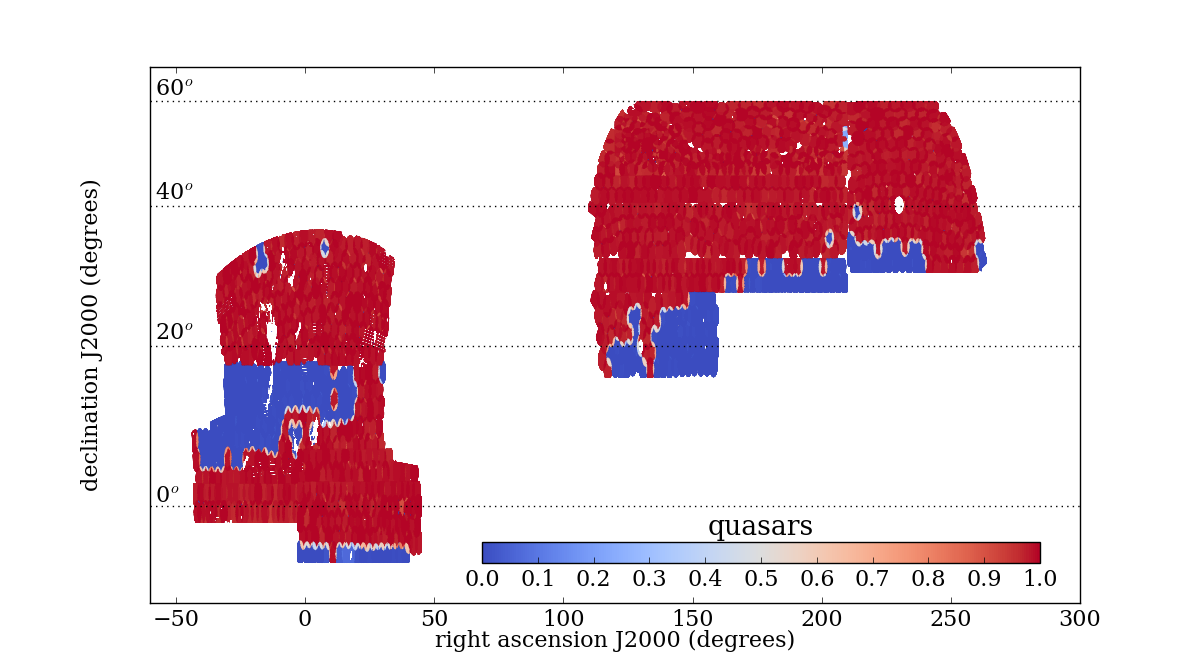}
 \includegraphics[width=3.25in]{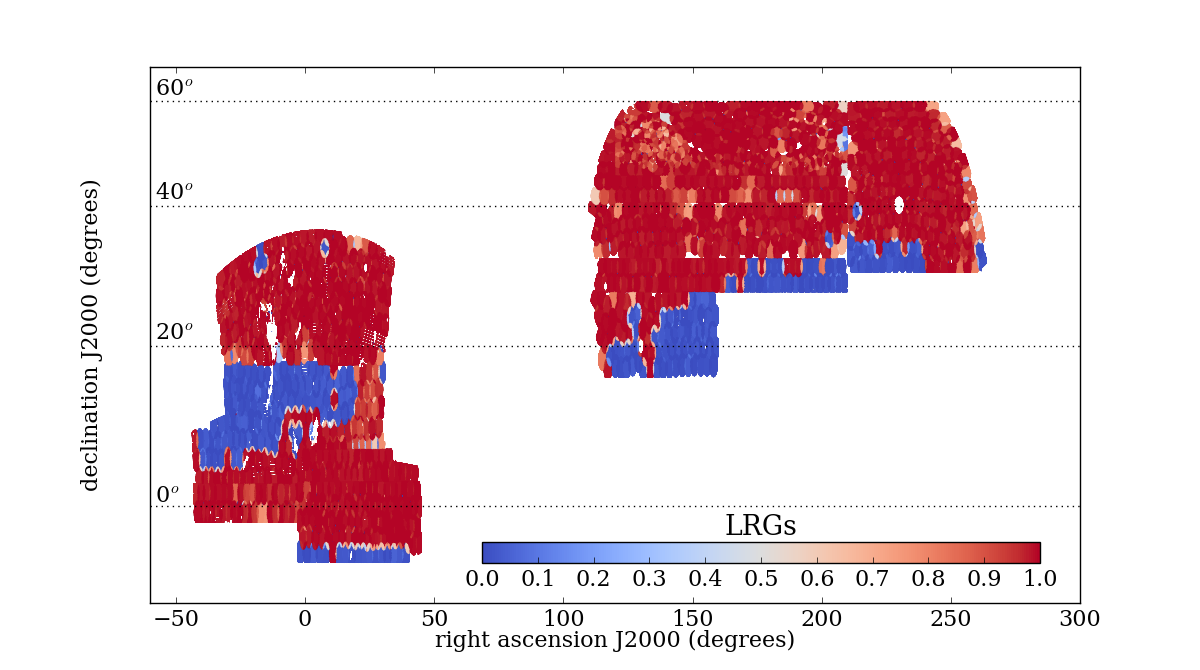}

  \caption{The footprint of eBOSS quasars (top) and LRGs (bottom). The colormap denotes the completeness. The dark blue areas were not observed by eBOSS (though the areas were tiled). }
  \label{fig:LRGquasarfoot}
\end{figure}

The completeness of the quasar sectors is shown in the top panel of Fig. \ref{fig:LRGquasarfoot}. One can see that the areas that had any plates observed are highly complete, but there were large areas that were tiled but not observed. Cutting to sectors that have $C_{\rm eBOSS} > 0.5$, the completeness of the quasar sample is 97.7 per cent. The statistics for the quasar sample are presented in Table \ref{tab:QSOstat}. One can observe that legacy redshifts make up more than one quarter of the total sample, with a higher percentage in the NGC than in the SGC (30 per cent compared to 22 per cent calculated as a fraction of $N_{\rm eff}$).

The completeness of the LRG sectors is shown in the bottom panel of Fig. \ref{fig:LRGquasarfoot}. One can see that it is nearly the same as that of the quasars, but has lower completeness. While the difference appears significant, the mean completeness when cutting to sectors that have $C_{\rm eBOSS} > 0.5$ is only 1 per cent less (96.7 per cent) than that of the quasar sample. The 480 deg$^2$ decrease in weighted area compared to the quasar sample is not apparent, since this is due almost entirely to the collision priority mask, which removes small holes of radius 62$^{\prime\prime}$. The statistics for the LRG sample are presented in Table \ref{tab:LRGstat}. 

\begin{figure}
\centering
\includegraphics[width=3.5in]{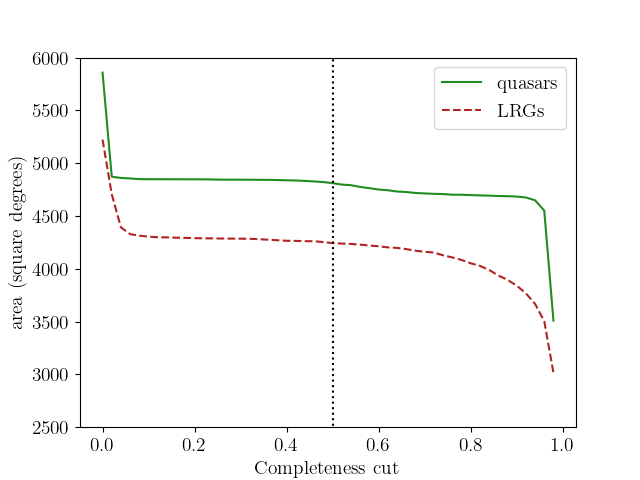}
  \caption{The area in the eBOSS footprint as a function of the completeness threshold, $C_{\rm eBOSS}$, for quasars and LRGs. The dotted line is at completeness of 0.5, which is the threshold applied to our clustering catalogs.}
  \label{fig:areavscomp}
\end{figure}

Fig. \ref{fig:areavscomp} displays the area in the eBOSS footprint greater than the completeness, $C_{\rm eBOSS}$, shown on the $x$-axis. The amount of area with $0.1 < C_{\rm eBOSS} < 0.7$ is only 135 deg$^2$ for the quasars and 144 deg$^2$ for the LRGs. The threshold applied to the eBOSS clustering catalogs is 0.5, which is shown with the dotted line. This matches the cuts applied to DR14 analyses. Just over 3,000 total LRG and quasar redshifts are removed by this cut, i.e., we lose less than 1 per cent of our eBOSS observations due this completeness cut.

We further track the redshift success rate per sector, $C_z$. This is given by
\begin{equation}
C_{z} = \frac{N_{\rm z,eboss} +N_{\rm badclass}+N_{\rm star} }{N_{\rm z,eboss}+N_{\rm cp} +N_{\rm badclass}+N_{\rm star} + N_{\rm zfail}}.
\end{equation}
We will apply a cut $C_{z} > 0.5$ to each sample. As shown in Table \ref{tab:mask}, this cut removes only an additional 3.4 deg$^2$ for the quasar sample and 1.6 deg$^2$ for the LRG sample. The change in footprint area as a function of this cut is quite small, as in the range $0 < C_z < 0.8$ it changes by only 7.5 deg$^2$ for quasars and by 17.1 deg$^2$ for the LRGs.  

\begin{figure}
\centering
\includegraphics[width=3.5in]{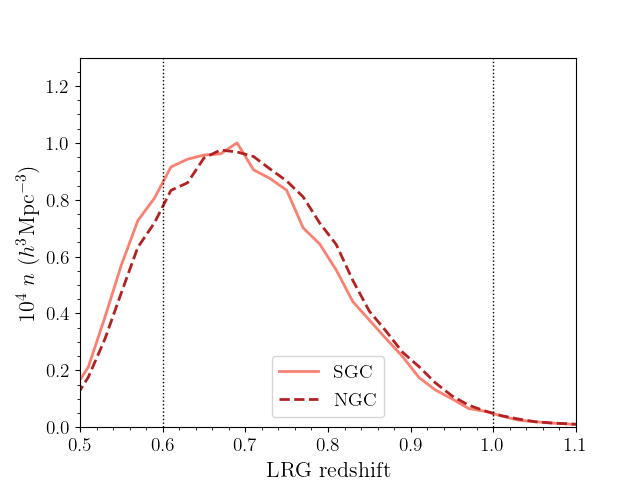}
\includegraphics[width=3.5in]{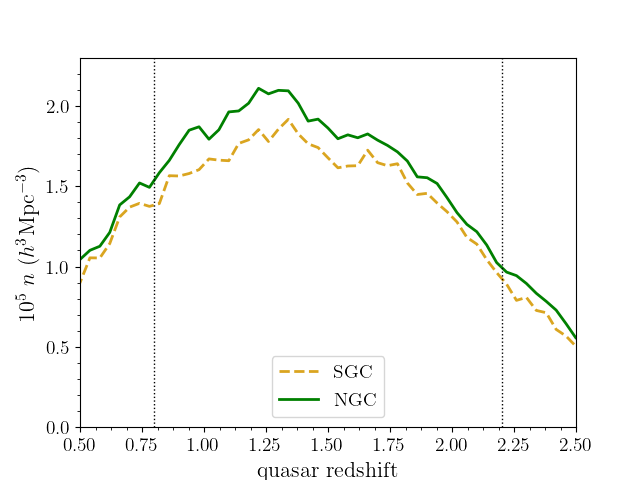}
  \caption{The number density of LRGs (top) and quasars (bottom) as a function of redshift, for the NGC and SGC. The vertical lines display the redshift cuts applied to create the clustering catalogs.}
  \label{fig:nzboth}
\end{figure}

Statistics for the quasars and LRGs after applying the completeness cuts are given in the bottom rows of Tables \ref{tab:QSOstat} and \ref{tab:LRGstat}. At this point, we also determine the $n(z)$ in the SGC and NGC for each sample. This is shown for the LRGs and quasars in Fig. \ref{fig:nzboth}. One can observe that the $n(z)$ for both samples are significantly different between the NGC and SGC. For the quasars, the difference is primarily a 10 per cent lower density in the SGC that is nearly constant with redshift. This is due to the difference in the mean depth between the two regions. The variations in density versus depth are explored in Section \ref{sec:imweight}. For the LRGs, the shapes of the $n(z)$ are not consistent. The specific differences between the NGC and SGC are not an issue for our samples, as the selection functions are estimated separately for the NGC and SGC. However, systematic variation in the shape of the $n(z)$ within either region is a systematic concern that we do not treat in our catalog construction\footnote{The overall variation in the $n(z)$ is accounted for with the weights determined in Section \ref{sec:imweight}} and was not found to be important in the LRG analysis \citep{BautistaDR16}. Strong variations in the shape of the $n(z)$ were found for the eBOSS ELG sample \citep{RaichoorELGcat} and found to be important to treat in the RSD analysis \citep{deMattiaELG,TamoneDR16}.

The $n(z)$ information is used to determine the \cite{FKP} weights for the sample
\begin{equation}
w_{\rm FKP} = 1/[1+n(z)P_0].
\end{equation}
We use $P_0 = 6000 ~({\rm Mpc}/h)^3$ for the quasars and $10,000~({\rm Mpc}/h)^3$ for the LRGs. These values match the amplitude of the power spectrum at $k\sim0.15h$Mpc$^{-1}$, which is the optimal choice for BAO analyses \citep{FB14}.

\subsection{Weights for imaging systematics}
We follow a similar approach to previous BOSS and eBOSS studies \citep{Ross12,Ross17,Ata,BautistaDR14LRG} and determine weights to correct for trends with properties of the imaging and Galactic foregrounds based on linear regression. The method is most similar to \cite{BautistaDR14LRG}. A multi-variate linear regression is performed, comparing {\sc healpix} \citep{healpix} maps of the projected sample density to those of imaging properties and Galactic foregrounds. The imaging conditions of SDSS are mapped at a  {\sc healpix} resolution\footnote{{\sc healpix} splits the sky into 12$N_{\rm side}^2$ equal area pixels.} $N_{\rm side}=512$. The map was created from a dense random sample that directly queried the SDSS imaging properties over the full eBOSS footprint. We will use maps of the depth in the $g$ band, the PSF size in the $i$ band, the sky background in the $i$ band, the airmass, and the \cite{SFD} Galactic extinction (E[B-V]). The particular choice of band is mostly arbitrary, as the SDSS imaging properties are highly correlated between bands. We use the same SDSS stellar density map, at $N_{\rm side}=256$ as used in previous analyses (e.g. \citealt{Ata,BautistaDR14LRG}). 

We fit for a different set of maps for the LRGs and quasars.  For each the spectroscopic completeness ($w_{\rm cp} w_{\rm noz}$) and $w_{\rm FKP}$ weights are applied along with the completeness cuts described in Section \ref{sec:comp} to create the maps. Further, the regression for each is performed separately for the NGC and SGC and the catalogs are cut to their target redshift range. For the LRGs, this is $0.6 < z < 1.0$ and for the quasars this is $0.8 < z < 2.2$. We regress against a given set of maps and, similar to the correction for redshift inefficiencies, we simply use the inverse of the fit as the weight ($w_{\rm sys}$) to apply to each object to correct for imaging systematics.  We define
\begin{equation}
w_{\rm sys} = \left[A_{\rm sys} + \vec{C}_{\rm sys}\cdot \vec{P}\right]^{-1},
\end{equation}
where $\vec{C}_{\rm sys}$ is the vector representing the coefficients fit to the set of maps with values $\vec{P}$ at the location of a given object.

\begin{table}
\centering
\caption{Coefficients for the linear regressions used to determine the values of weights to correct for systematic trends with characteristics of the imaging data. The regressions are performed separately for the NGC and SGC data. Both LRGs and quasars are regressed against Galactic extinction ($E[B-V]$), $i$-band sky background (sky$_i$) and the PSF size in the $i$-band (PSF$_i$). The extinction-corrected $g$-band depth (depth$_g$) is additionally included in the regression against quasar density. For the LRG regression, stellar density ($\delta_{\rm star}$) is included. $A_{\rm sys}$ is the constant in the linear regression.}
\begin{tabular}{lcccccc}
\hline
Sample	& $\delta_{\rm star}$ & depth$_g$ & $E[B-V]$ & sky$_i$ & PSF$_i$ & $A_{\rm sys}$ \\\hline\hline
NGC quasars & - & 0.11 & -0.14 & -0.038 & -0.095 & 0.030 \\
SGC quasars & - & 0.25 & -0.12 & -0.075 & -0.12 & -0.034 \\
NGC LRGs & -0.25 & - & -0.17 & 0.17 & -0.062 & 0.056 \\
SGC LRGs & -0.48 & - & -0.042 & 0.096 & -0.075 & 0.098 \\

\label{tab:syscoef}
\end{tabular}
\end{table}

\begin{figure}
\centering
\includegraphics[width=3.5in]{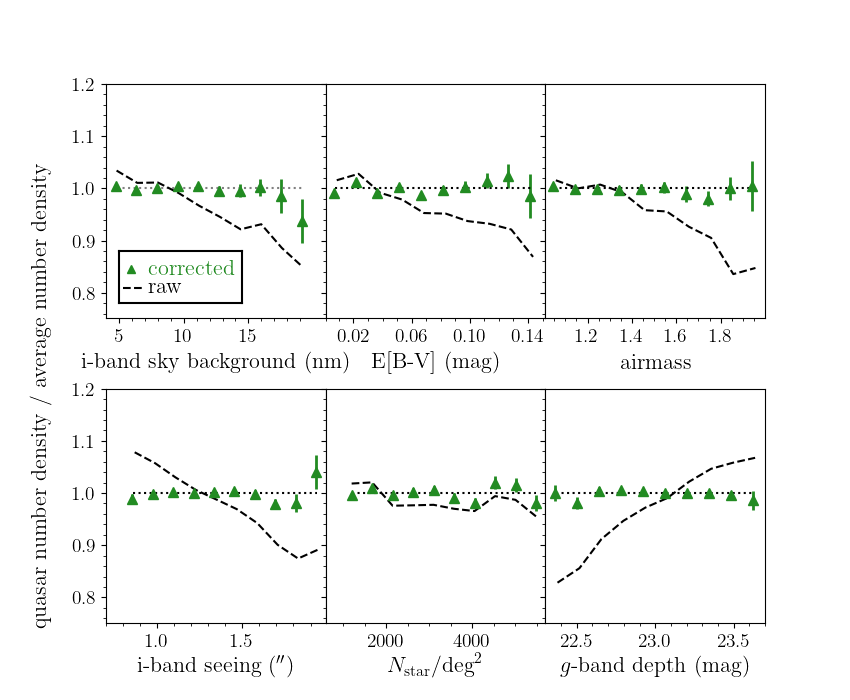}
  \caption{Fluctuations in projected quasar density as a function of various image properties and Galactic foregrounds, combining NGC and SGC results (but normalizing them separately). The dashed curves show the result before weights for $g$-band depth and E(B-V) are applied. }
  \label{fig:qsoimsys}
\end{figure}

For the quasars, we use the maps that capture the SDSS imaging depth in the $g$-band and Galactic extinction, as was done in the DR14 analysis \citep{Ata}.  We further include maps of the sky background and seeing. 
 The coefficients determined from the regressions are included in Table~\ref{tab:syscoef}.
Fig. \ref{fig:qsoimsys} displays fluctuations in the projected quasar density as a function of the imaging properties and Galactic foregrounds considered in our analysis. We have combined NGC and SGC for these results, but normalized them separately. (Given that the five SDSS imaging bands were observed nearly simultaneously via drift scan, the specific bands are nearly perfectly correlated.) One can observe that strong trends with all maps are greatly reduced after the weights are applied. For instance, a strong trend with airmass is removed, despite us not including that map in the regression. This is due to the fact that airmass is one contributing factor to the depth. The $\chi^2$ for the null test after applying the systematic weights (ignoring any covariance) are 6, 34, 5, 9, 20, 5 left-to-right, top-to-bottom. Four maps were used in the regression and 10 measurement bins are used in this test, so the total $\chi^2$/dof is 79/55. The strongest residual is for the Galactic extinction (E[B-V]), despite the fact it was one of the maps used in the regression. The impact of any residual uncertainty on RSD or BAO analyses with respect to these imaging weights is studied further in \cite{HouDR16,NeveuxDR16}. Primordial non-Gaussianity studies are particularly sensitive to the large-scale power that can be introduced by these kind of systematic fluctuations \citep{Huterer13,Ross13}.  The impact of residuals and whether further cleaning for such primordial non-Gaussianity analyses is possible is being investigated by \cite{RezaiefNL,MuellerfNL}.

\begin{figure}
\centering
\includegraphics[width=3.5in]{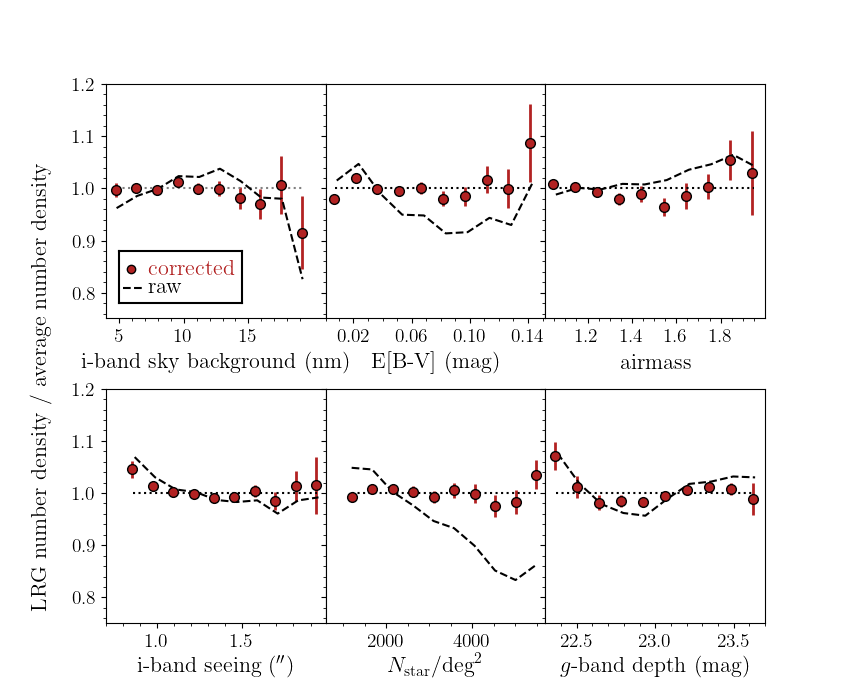}
  \caption{Fluctuations in projected LRG density as a function of various image properties and Galactic foregrounds, combining NGC and SGC results (but normalizing them separately). The dashed curves show the result before weights for stellar density and E(B-V) are applied. }
  \label{fig:lrgimsys}
\end{figure}

For the LRGs, the systematic correlation is strongest with stellar density (consistent with \citealt{BautistaDR14LRG}).  We additionally include the Galactic extinction, sky background, and seeing maps in the regression. The coefficients determined from the regressions are included in Table~\ref{tab:syscoef}. Fig. \ref{fig:lrgimsys} displays fluctuations in the LRG density before (dashed curves) and after (points with error-bars) the weights are applied. The $\chi^2$ for the null test after applying the systematic weights (ignoring any covariance) are 6, 30, 14, 13, 8, 20 left-to-right, top-to-bottom. The total $\chi^2$/dof is 91/55.  Similar to the results for the quasars, the strongest residual is for the Galactic extinction (E[B-V]), despite the fact it was one of the maps used in the regression. The impact of any residual uncertainty on RSD or BAO analyses with respect to these imaging weights is studied further in \cite{GilMarinDR16,BautistaDR16}.

\label{sec:imweight}

\subsection{Assigning radial selection function to randoms}

In order to assign redshifts and relevant information to the random catalogs, we follow the same procedure as in previous BOSS and eBOSS analyses and randomly select redshifts from the relevant observed sample. A difference for the LRG and quasar catalogs, however, is that all of the columns that are used for the data sample are also used for the random catalog. 

As a first step, the $C_{\rm eBOSS}$ information is copied to the $w_{\rm sys}$ column of the random catalog. Then, for each ra,dec in the random catalog, a random LRG/quasar is selected. Its redshift, $w_{\rm cp}$, $w_{\rm noz}$, and $w_{\rm FKP}$ are assigned to the row. Its $w_{\rm sys}$ is multiplied by the existing $w_{\rm sys}$ to provide the total value for this quantity.

Treating the randoms in this fashion means that the random catalogs are processed in exactly the same way as the data file in order to calculate any statistics. Namely, the total contribution for any data/random point is given by

\begin{equation}
w_{\rm tot} = w_{\rm sys}w_{\rm cp}w_{\rm noz}.
\end{equation}

We further recommend multiplying both the data and random $w_{\rm tot}$ by $w_{\rm FKP}$ in order to produce more optimally weighted clustering statistics. 

\subsection{Combining eBOSS LRGs and BOSS CMASS}
\begin{figure}
\centering
\includegraphics[width=3.5in]{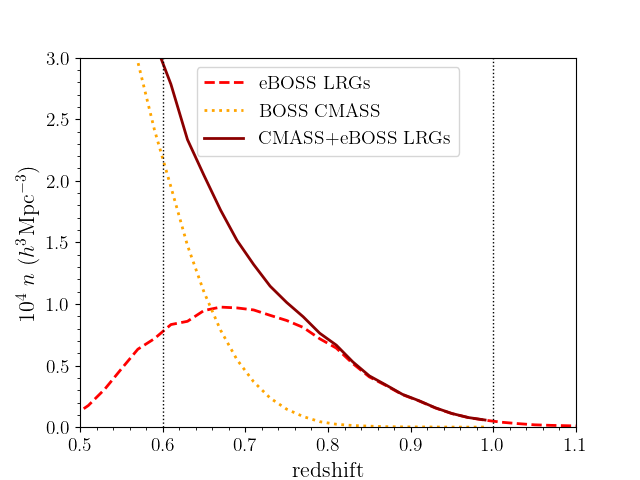}
  \caption{The number density of BOSS CMASS, eBOSS LRGs, and their combination, in the NGC region. The combined sample is used within the eBOSS footprint, while CMASS only is used outside of the eBOSS footprint.}
  \label{fig:nzce}
\end{figure}

We combine eBOSS LRGs and BOSS CMASS galaxies with $z > 0.6$ in order to create one sample to be used for cosmological analysis. For CMASS, we make few alterations to the sample defined in \cite{Reid16}. We first cut the sample to $z>0.6$. We then enforce that the ratio of weighted randoms to weighted data is the same for both CMASS and eBOSS LRGs. We then determine which CMASS galaxies are within the eBOSS footprint, using {\sc Mangle} to match to the eBOSS sectors. We assume all eBOSS LRGs are within the CMASS footprint. Within the eBOSS sectors, the $n(z)$ is recalculated by adding the LRG and CMASS $n(z)$. The $n(z)$ for each sample is shown in Fig. \ref{fig:nzce}. This new $n(z)$ is used to recalculate the $w_{\rm FKP}$ within the eBOSS region. Outside of the eBOSS region, the CMASS catalogs remain the same. The definition of the spectroscopic completion weights is different in CMASS. Thus, for convenience, we produce a $w_{\rm tot}$ column\footnote{It is named `WEIGHT\_ALL\_NOFKP'.}, applying the appropriate algorithm to each sample.

\begin{table*}
\centering
\caption{Some statistics for the area and number of eBOSS and BOSS galaxies that enter the combined LRG$+$ CMASS sample with $0.6 < z < 1.0$. For each division of the sample, we quote the $N_{\rm eff}$, which is the sum of number of galaxies weighted by the close-pair and redshift failure weights.}
\begin{tabular}{lrrrrrrrrr}
\hline
 & &  BOSS only & & & Overlap & & & Combined &  \\
\hline
	& SGC & NGC~~ & Total~~ & SGC & NGC~~ & Total~~ & SGC & NGC~~ & Total\\
\hline\hline
Area (deg$^2$)  & 884 & 4,368~~ & 5,251 &  1,676 & 2,566~~ & 4,242 & 2,560 & 6,934~~ & 9,493 \\
$N_{\rm eff}$ eBOSS & 0 & 0~~ & 0  & 71,427 & 113,868~~ & 185,295 & 71,427 & 113,868~~ & 185,295  \\
$N_{z}$ eBOSS & 0 & 0~~ & 0 & 67,316 & 107,500~~ & 174,816 & 67,316 & 107,500~~ & 174,816\\
$N_{\rm eff}$ BOSS  & 16,645 & 95,247~~ & 111,892 & 41,753 & 63,112~~ & 104,865 & 58,398 & 158,358~~ & 216,756 \\
$N_{z}$ BOSS & 15,495 & 88,952~~ & 104,447& 38,906 & 59,289~~ & 98,195 & 54,401 & 148,241~~ & 202,642  \\
$N_{\rm eff}$ BOSS$+$eBOSS & 16,645 & 95,247~~ & 111,892 & 113,180 & 176,980~~ & 290,160 & 129,825 & 272,226~~ & 402,052  \\
$N_{z}$ BOSS$+$eBOSS & 15,495 & 88,952~~ & 104,447 & 106,222 & 166,789~~ & 273,011 & 121,717 & 255,741~~ & 377,458  \\
\hline\hline
\label{tab:LRGCMASS}
\end{tabular}
\end{table*}

Some statistics for the combined LRG$+$CMASS sample are given in Table \ref{tab:LRGCMASS}. Overall, BOSS CMASS galaxies make up slightly more than half of the total sample and the area they occupy is more than twice that of eBOSS LRGs. One can see the footprints of the eBOSS and CMASS areas in Fig. \ref{fig:DR16obs}. The majority (65 per cent) of the CMASS SGC was observed by the eBOSS LRG program, while eBOSS covered 37 per cent of the NGC CMASS area. Thus, the fraction of the sample that is comprised of eBOSS LRGs is different in each region, 55 compared to 43 per cent. The projected angular number density of galaxies with $0.6 < z < 1.0$ is nearly twice as high for the eBOSS LRGs compared to CMASS (44 deg$^{-2}$ compared to 23 deg$^{-2}$). 

The combined LRG$+$CMASS sample is run through the reconstruction algorithm described in \cite{burden14,BautistaDR14LRG}. This moves overdensities back along an estimate of the vector of linear motion, removing some of the non-linear dispersion signal from the data. This, in turn, sharpens the BAO peak and thus increases the precision of BAO-based distance-redshift measurement \citep{Eisenstein07rec}. In our companion papers, reconstruction is only applied for BAO measurements and not those that use the RSD signal. The LRG catalog differs from the quasar catalog in this regard, as reconstruction was not applied to the quasars due to the lower number density. We provide LRG catalogs with and without reconstruction applied. Further details on how the reconstruction algorithm was applied are presented in \cite{BautistaDR16}.

\section{Conclusions}\label{sec:conclusion}

We have described the creation of LSS catalogs for the eBOSS DR16 LRG and quasar samples. The LRG catalog is combined with BOSS CMASS and the combined catalog contains 377,458 galaxies with $0.6 < z < 1.0$ intended for cosmological analysis. Likewise, our catalog contains 343,708 quasars with $0.8 < z < 2.2$. For each sample, there is a random sample that is at least 40 times more dense and approximates the respective three dimensional (ra, dec, redshift) selection function. Weights are provided for both the data and random samples in order to ensure the randoms do match the selection function and optimize the signal-to-noise of the clustering measurements. These catalogs are available to the public at https://data.sdss.org/sas/dr16/eboss/lss/catalogs/DR16/.  

Our descriptions allow our results to be reproduced. They further allow the companion analyses to study systematic uncertainties imparted during any part of the process. Thus, in addition to analyzing the two-point statistics of the catalogs reported here, companion papers use this information to simulate the LRG and quasar samples and demonstrate that systematic uncertainties are a sub-dominant component to the LRG and quasar cosmological results. In particular:
\begin{itemize}
\item The uncertainties on quasar redshifts are characterized in Section \ref{sec:qsoz}, based on inputs from \cite{LykeDR16} and including the rate of catastrophic failures. In \cite{SmithDR16} these results are used to create simulations with varying assumptions  on the redshift error distribution. \cite{NeveuxDR16,HouDR16} apply their analyses to these simulations in order to quantify the level of systematic uncertainty introduced from these redshift uncertainties.
\item The completeness map and $n(z)$ (Section~\ref{sec:comp}) are used in \cite{ZhaoDR16} to produce mock LRG and quasar surveys that are used for covariance matrix estimation.
\item \cite{ZhaoDR16} further approximate the process used to determine weights for spectroscopic completeness (described in Section \ref{sec:specweight}) and imaging systematics (described in Section \ref{sec:imweight}). This allows for the estimation of systematic uncertainty related to these corrections.
The results for LRG clustering measurements are in \cite{BautistaDR16,GilMarinDR16} and for quasar clustering measurements in \cite{NeveuxDR16,HouDR16}. In all cases, the total observational systematic uncertainty is found to be sub-dominant compared to the statistical uncertainty.
\end{itemize}

In addition to the LRG and LSS quasar analyses, eBOSS DR16 includes studies of ELG clustering and the Ly-$\alpha$ forest. The ELG catalogs are presented in \citet{RaichoorELGcat}, and analyzed in \citet{TamoneDR16}, \citet{deMattiaELG} and \citet{AlamDR16}. The DR16 Ly-$\alpha$ sample is presented and analyzed in \citet{dumasdesbourbouxDR16}. The results from all of the eBOSS tracers are used in \cite{DR16cosmo} in order to update our understanding of cosmology.

The public release of these catalogs marks the end of the eBOSS experiment.
After two decades of cosmology, eBOSS represents the conclusion
of LSS surveys performed by the Sloan telescope. DESI \citep{DESI,DESIin} is the spectroscopic cosmology program that is
the natural successor to eBOSS.
DESI achieved first light in late 2019 and will have approximately 20 times the power of the Sloan telescope + BOSS spectrograph
for conducting galaxy surveys. We expect significant research will be required to update the methods for catalog creation
presented here, in particular to model focal plane incompleteness and $n(z)$ variation, in order to maintain systematic uncertainties that are below the sub-percent statistical precision that is expected
from that program. 

\section*{acknowledgements}
AJR is grateful for support from the Ohio State University Center for Cosmology and Particle Physics. 
SA is supported by the European Research Council through the COSFORM Research Grant (\#670193)
BL and ADM were supported by the U.S.Department of Energy, Office of Science, Office of High Energy Physics, under Award Number DE-SC0019022.
Authors acknowledge support from the ANR eBOSS project (ANR-16-CE31-0021) of the French National Research Agency.
G.R. is supported by the National Research Foundation of Korea (NRF) through Grants No. 2017R1E1A1A01077508 and No. 2020R1A2C1005655 funded by the Korean Ministry of Education, Science and Technology (MoEST), and by the faculty research fund of Sejong University.

Funding for the Sloan Digital Sky Survey IV has been provided by the Alfred P. Sloan Foundation, the U.S. Department of Energy Office of Science, and the Participating Institutions. SDSS-IV acknowledges
support and resources from the Center for High-Performance Computing at
the University of Utah. The SDSS web site is www.sdss.org. This work made use of the facilities and staff of the UK Sciama High Performance Computing cluster supported by the ICG, SEPNet and the University of Portsmouth.
In addition, this research relied on resources provided to the eBOSS
Collaboration by the National Energy Research Scientific Computing
Center (NERSC).  NERSC is a U.S. Department of Energy Office of Science
User Facility operated under Contract No. DE-AC02-05CH11231.

SDSS-IV is managed by the Astrophysical Research Consortium for the 
Participating Institutions of the SDSS Collaboration including the 
Brazilian Participation Group, the Carnegie Institution for Science, 
Carnegie Mellon University, the Chilean Participation Group, the French Participation Group, Harvard-Smithsonian Center for Astrophysics, 
Instituto de Astrof\'isica de Canarias, The Johns Hopkins University, Kavli Institute for the Physics and Mathematics of the Universe (IPMU) / 
University of Tokyo, the Korean Participation Group, Lawrence Berkeley National Laboratory, 
Leibniz Institut f\"ur Astrophysik Potsdam (AIP),  
Max-Planck-Institut f\"ur Astronomie (MPIA Heidelberg), 
Max-Planck-Institut f\"ur Astrophysik (MPA Garching), 
Max-Planck-Institut f\"ur Extraterrestrische Physik (MPE), 
National Astronomical Observatories of China, New Mexico State University, 
New York University, University of Notre Dame, 
Observat\'ario Nacional / MCTI, The Ohio State University, 
Pennsylvania State University, Shanghai Astronomical Observatory, 
United Kingdom Participation Group,
Universidad Nacional Aut\'onoma de M\'exico, University of Arizona, 
University of Colorado Boulder, University of Oxford, University of Portsmouth, 
University of Utah, University of Virginia, University of Washington, University of Wisconsin, 
Vanderbilt University, and Yale University.

\section*{Data Availability}
The data described in this paper are available to the public at https://data.sdss.org/sas/dr16/eboss/lss/catalogs/DR16/ and are described at https://www.sdss.org/dr16/spectro/lss/.

\label{lastpage}

\end{document}